\documentclass[12pt]{article}
\usepackage[T1]{fontenc}         
\usepackage{amsmath,bbm,amssymb}
\usepackage{amsthm,amscd}
\usepackage{graphicx}            
\usepackage{comment}
\usepackage{longtable, lscape}
\usepackage{rotating}
\usepackage{algorithm}
\usepackage{xurl}
\usepackage[hidelinks,breaklinks=true]{hyperref}
\usepackage{multirow}
\usepackage{subfig}
\usepackage{color}
\usepackage{natbib}
\usepackage{url}             
\usepackage{placeins}            
\usepackage{booktabs}            
\usepackage{siunitx}             
\usepackage{dcolumn}             
\usepackage{threeparttable}      
\usepackage{array}               
\graphicspath{{results/}{./}}

\usepackage[title]{appendix}

\newcommand{\R}{\mathbb{R}}

\newcommand{\Norm}{\mathcal{N}}
\newcommand{\T}{^{\!\top}}

\newcommand{\bze}{\mathbf{0}}

\newcommand{\by}{\mathbf{y}}
\newcommand{\ba}{\mathbf{a}}
\newcommand{\bu}{\mathbf{u}}

\newcommand{\bB}{\mathbf{B}}
\newcommand{\bX}{\mathbf{X}}
\newcommand{\bY}{\mathbf{Y}}
\newcommand{\bH}{\mathbf{H}}
\newcommand{\bI}{\mathbf{I}}
\newcommand{\bER}{\mathbf{E}}

\newcommand{\bSig}{\boldsymbol{\Sigma}}

\newcommand{\bbeta}{\boldsymbol{\beta}}

\newcommand{\SH}{\mathbf{S}_{\mathrm{H}}}
\newcommand{\SE}{\mathbf{S}_{\mathrm{E}}}

\DeclareMathOperator{\tr}{tr}

\begin{document}
\baselineskip=22pt
\vskip 20pt
\begin{center}
{\Large \bf An Experimental Design Approach to Evaluating Agentic AI's Autonomous Model Discovery 
}\\
\vskip 5pt
Hao He\textsuperscript{1}, Xueying Liu\textsuperscript{2}, Chris J. Kuhlman\textsuperscript{3}, and Xinwei Deng\textsuperscript{1}\footnote{Address for Correspondence: Xinwei Deng, Professor, Department of Statistics, Virginia Tech, Blacksburg, VA 24061 (Email: xdeng@vt.edu).}\\
\textsuperscript{1}Department of Statistics, Virginia Tech \\ \textsuperscript{2}Department of Statistical Science, Baylor University \\ \textsuperscript{3}Advanced Research Computing, Virginia Tech\\
\end{center}
\vskip 5pt
\begin{abstract}

Large language model coding agents increasingly perform open-ended data modeling and analysis. These agents are stochastic and adaptive, and therefore their autonomous model discovery behavior cannot be adequately characterized by a single benchmark run. In this work, we propose an experimental design and analysis framework for systematically evaluating this discovery process, quantifying its variability, and identifying important factors. The proposed framework treats these agents as stochastic model-discovery operators, which map task-specific discovery data and an optimization target to a fitted model. Specifically, we investigate two such operators, Codex and Claude Code, under controlled experimental factors including agent’s reasoning effort, task, optimization metric, and composition of training data. For each agent-task-metric combination, regression models and inference are conducted for multiple responses such as output quality, dollar cost, wall-clock time, and process complexity. Furthermore, we develop a utility-aligned canonical decomposition to characterize the dominant direction of the reasoning-effort effect and to assess whether that direction aligns with a performance-cost utility direction. The proposed framework is demonstrated on a testbed of networked word-forming games with insightful findings on reasoning effort with respect to cost and process complexity.
\end{abstract}

\vskip 8pt
\section{Introduction}
\label{sec:intro}

Agentic AI and large language coding agents now carry out the technical work of data modeling and
analysis without human intervention. These agents are stochastic, adaptive, sequential decision-making
systems, and their autonomous discovery behavior cannot be adequately evaluated by single-run
benchmarks. How such an agent's behavior changes as its operating controls vary, and what that
behavior costs, is not yet well understood. Our motivating case
study is from the networked group anagram game of forming words, where a small team of human
players connects through a communication network and each player starts with a few letters
of the alphabet. Over a five-minute session, every player repeatedly chooses among four
actions: form a word from the letters in hand, request a letter from a network neighbor,
reply to a request by sharing a letter, or sit idle.
The team's earnings grow with the number of words the group forms and are split evenly among
the members, so the payoff structure rewards cooperation. For this simple-rule game, a
player's decisions are inductive and interconnected: they depend on the current inventory of
letters, on pending requests, on what neighbors have just done, and on the shared incentive
to produce words as a group. Throughout the game, one can trace out a discrete-time
behavioral process in which individual players' actions interact through the
network to produce group-level behavior. The data obtained from each
session (i.e. game) are therefore sequences of interdependent actions on a network rather
than independent records. Every observation is one player's discrete choice at one time step,
coupled to that player's history, to neighbors' concurrent behaviors, and to the group's
shared payoff.

\begin{figure}[!t]
\centering
\includegraphics[width=\textwidth]{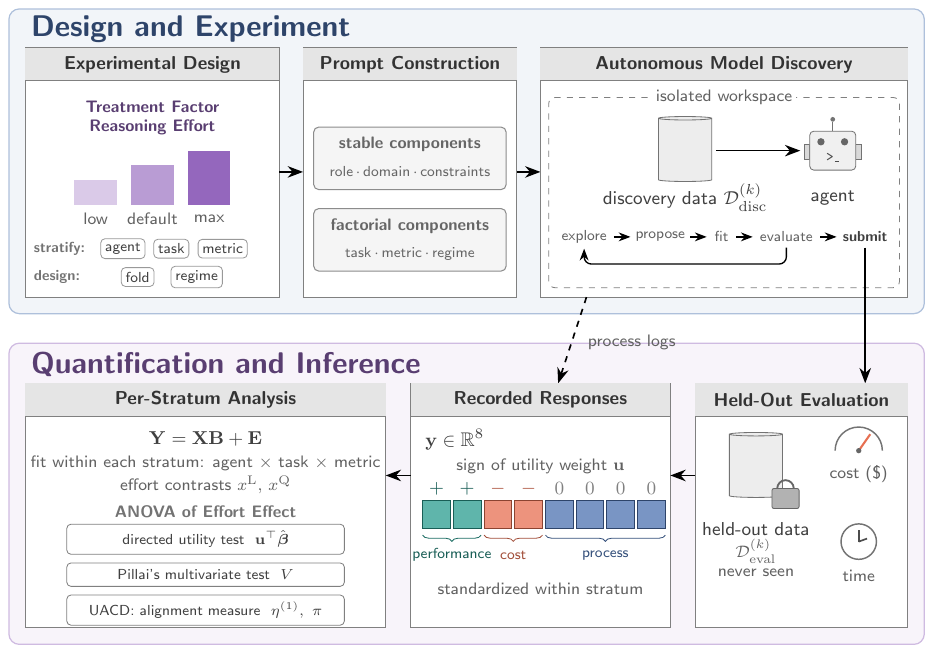}
\caption{Illustration of the experiment and analysis pipeline.}
\label{fig:pipeline}
\end{figure}

Modeling such data is very useful to study cooperation in small groups for behavioral and
data scientists. In the literature, modeling efforts have been made in the social-network
analysis and simulation communities: mechanistic and data-driven agent-based models of
player behavior \citep{ren2018generative, cedenomieles2019anagram}, uncertainty
quantification and calibration for those models \citep{hu2024uncertainty}, and recent
extensions to digital twins and to LLM-assisted model improvement
\citep{he2024digitaltwin, he2025model}. Note that modeling such temporal, network-coupled
action data is more than routine regression or classification models. It
requires models that reproduce group behaviors as well as individual choices. Moreover, modeling these data requires multiple iterations with researchers proposing an underlying
mechanism, conducting feature engineering, exploratory analysis, data modeling, performance
evaluation, and modeling modification.

The modeling effort of proposal-modeling-evaluation loop is exactly the kind of
open-ended technical work that large language model (LLM) coding agents now perform
\citep{jimenez2024swebench}. We call such an agent a
\emph{model-discovery operator}, which can read and explore data, construct features, choose
a model class, write and run code, decide what to return, and produce a final report
including script and evaluation. Acting as a modeler, this operator is
stochastic and its output is uncertain. The prospect is appealing, since cycles of the
workflow become an automated run without expert intervention. Yet an AI agent is not a fixed
machine. It contains multiple control configurations the user needs to set, such as which
LLM agent and model version to use, how much reasoning effort to request, and how much
training data to provide. For example, more requested effort might buy better features, more
careful validation, and stronger held-out performance. It might also produce longer code,
require more tokens and higher cost. Whether added effort improves the performance or only raises its cost is an
empirical question. The same holds for how quality, cost, and process move together as the
controls change.

To address these challenges, we propose an experimental design and analysis framework that
systematically probes the AI model-discovery operator through data from networked
anagram games, measuring the variability of its discovery process and identifying the factors
that affect it. The proposed method examines the AI model-discovery operator with a controlled
factorial experiment over its configuration controls under different tasks.
Figure~\ref{fig:pipeline} gives an overview of the study, from the factorial design through
the agent's autonomous discovery run to the held-out scoring and the per-stratum multivariate
analysis.

Specifically, we consider two discovery tasks on the data. The first task is predictive, from
a player's recent history and local network context, on forecasting the player's next action.
The second is generative, from an agent-based simulator perspective, on generating the whole
sequences of actions of a session. For each task, the proposed experimental design framework
considers two commercial coding agents, the primary metric each task is
scored on, three levels of requested reasoning effort as the main treatment factor,
the cross-validation fold held out for scoring, and training-data regimes. By
recording the complete trace of every run, we can collect comprehensive
information such as the candidate models the AI agent explored, the tokens and dollars it
spent, the time it took, and the script it finally submitted. Thus we can treat performance,
monetary and wall-clock cost, and process complexity as the joint response outcomes for
analyzing how the agent's controls, mainly the requested reasoning effort,
move them jointly.

Furthermore, the proposed analysis framework considers a Utility-Aligned Canonical
Decomposition (UACD) to identify the dominant direction in which effort moves
the response vector. It can pair a pre-specified utility contrast with the UACD. It provides
useful insights on whether the operator's effort-induced movement is aligned with the
pre-specified utility direction.

The proposed framework contributes a scientific-discovery testbed built on networked anagram
experiments. Note that our design and analysis framework is aimed at the AI model-discovery operator rather than
the anagram models it produces. It can yield operational guidance on how an agent's controls
move quality, cost, and process jointly. A documented dataset and a reproducible evaluation pipeline are publicly
available at \href{https://github.com/haohe13/starf-anagram-game-model-discovery-data-collection}{the project's GitHub repository}. \footnote{
\texttt{https://github.com/haohe13/starf-anagram-game-model-discovery-data-collection}
}

The remainder of the paper is organized as follows. Section~\ref{sec:rw} reviews related work. Section~\ref{sec:method} describes the data and the two discovery tasks, the model-discovery operators, the factorial design, the agents' optimization target and experiment outcomes. Section~\ref{sec:analysis} presents the multivariate analysis framework and the utility-aligned canonical decomposition. Section~\ref{sec:results} reports the empirical results. Section~\ref{sec:discussion} summarizes the findings and discusses limitations and future work. Appendices~\ref{sec:impl} and~\ref{sec:app-extra} give implementation details and additional exploratory results.

\section{Literature Review}
\label{sec:rw}

There is a body of work treating language models as stand-ins for human
respondents and as components of simulated populations. For example,
\citet{argyle2023outofone} show that conditioning a model on demographic and attitudinal
profiles reproduces the response distributions of human subpopulations.
\citet{aher2023simulating} consider LLMs as simulated participants and recover several
classic human-subject findings. Other works also use LLMs as stand-ins for a
simulated economic agent \citep{horton2023economic} and human agents
\citep{park2023generative}. \citet{gao2024survey} provide a review
of LLM-powered agent-based modeling and simulation. However, these works
focus largely on whether a model can behave like a person, and not much on
evaluating how an automated agent builds such a model.

There are also works that use LLMs to automate model search and parts of the data-analysis workflow \citep{romeraparedes2024funsearch, lu2024aiscientist, manning2024automated}. Closest to statistical model discovery, \citet{li2024automated} propose a language-model loop that alternates between proposing probabilistic programs and critiquing their fit, producing models comparable to expert-designed ones without a hand-built search procedure. Benchmarks have also been developed. For example, MLAgentBench
\citep{huang2024mlagentbench} considers evaluating agents on machine-learning
experimentation, extending the automated machine learning literature
\citep{thornton2013autoweka, feurer2015autosklearn}. The agentic search runs in open-ended
natural language and code with underlying goals: choosing a model family, constructing
features, fitting models, and deciding what to report. In this work, we
consider how those choices respond to an agent's controls, which is not
considered in the aforementioned research on benchmarks. Closest to our setting,
\citet{he2025model} use an LLM to improve behavioral models of the networked anagram
experiments. That work produced a single improved model, whereas here we focus on
characterizing the AI agent as a stochastic operator across many controlled runs.

How to evaluate such a stochastic model-discovery operator is of great
interest. There are different benchmarks \citep{chen2021humaneval, srivastava2023bigbench,
jimenez2024swebench} and different evaluation criteria \citep{liang2023helm, chiang2024arena,
zheng2023judging}. Our design and analysis methods consider holding the task fixed and
modeling how one agent's quality, cost, and process respond to its controls, treating cost
and process as outcomes, not diagnostics.

The design of experiments has been commonly used to investigate
complex systems \citep{montgomery2017doe, wu2011experiments}. Classical
experimental design and analysis often involve meticulous data-collection
planning with treatments from multiple important factors and the corresponding performance
measures. Separately, cost has begun to enter language-model evaluation explicitly. FrugalGPT
\citep{chen2023frugalgpt} studies cost-quality trade-offs and large-scale usage studies
\citep{aubakirova2025state} document the price and token economics of deployed models. To our
knowledge, no prior work combines these elements for an agentic operator. The existing
methods often measure whether an agent succeeds but they do not treat the agent itself as the
object of a designed experiment. Nor have cost-aware and
multivariate analyses been applied to agentic model discovery. Our work fills
this gap by treating an LLM coding agent as a factorial-design object whose performance, cost,
and process are joint responses to its controls.
\section{Experimental Design Framework}
\label{sec:method}

We first describe the data and the two model-discovery tasks and introduce the LLM
coding agents we study as model-discovery operators (Section~\ref{sec:tasks}).
Section~\ref{sec:design} presents the full factorial design over six factors, and
Section~\ref{sec:metrics} defines the target metrics and the eight outcomes that form the
multivariate response analyzed in Section~\ref{sec:analysis}.

\subsection{Data, Discovery Tasks, and Model-Discovery Operators}
\label{sec:tasks}

\paragraph{Data.}
All tasks use a fixed dataset from a networked group-anagram experiment
\citep{cedenomieles2019anagram,ren2018generative,he2025model}. Players are arranged on a communication
network, and each has a limited inventory of letters. The game lasts $t_{\max}=300$ seconds. At each time $t$ a player takes one of four actions:
forming a word the current letter inventory allows, requesting letters from a neighbor, replying to a
neighbor's request, or staying idle. The action of sharing duplicates a letter rather than
transferring it, so helping a neighbor never shrinks the helper's own inventory. We encode the
action of player $v$ at time $t$ as $a_{v,t}\in\{1,2,3,4\}$, with $1$ for
staying idle, $2$ for replying to a neighbor's request, $3$ for requesting a letter, and $4$
for forming a word. The data comprise $28$ sessions and $209$ players, giving $62{,}700$
player-time records. Twenty-four sessions use a closed-ring network, and the remaining four are \emph{partial-path} sessions in which the ring is opened into a path, leaving the two end players in each with a single neighbor rather than two (eight such players across the four sessions). The data have two parts: a relational part, the network, and a temporal part, the per-player action sequences. We split the whole data at the session level into three cross-validation folds $\mathcal{F}_1,\mathcal{F}_2,\mathcal{F}_3$, each having $9$, $9$, and $10$ sessions, respectively. For an experimental run with fold $k$ held out, we denote $\mathcal{D}_{\mathrm{eval}}^{(k)}$ as the held-out set, which will be used only for evaluation scoring and never seen during discovery. We denote $\mathcal{D}_{\mathrm{disc}}^{(k)}$ as the discovery data the agent learns from.

\paragraph{Tasks.}
We define two model-discovery tasks on these data, one predictive and one
generative. Task~1 (\emph{timestep}) is four-class next-action prediction. For each player $v$
at time $t$, the agent's submitted model predicts a probability vector $\widehat
p(a_{v,t+1}=j\mid h_{vt})$, $j\in\{1,2,3,4\}$, over the player's next action $a_{v,t+1}$, using
the player's history $h_{vt}$ through time $t$ and the player's network context. Task~2 (\emph{ABM}, for agent-based model) is generative. The agentic AI builds an agent-based simulator of a session, and the evaluation pipeline runs it forward
on held-out sessions and compares the result with the real held-out trajectories. One can see that the two tasks probe different notions of adequacy: a next-action predictive model can score well on individual next-action predictions yet miss the emergent dynamics of a full game, whereas an ABM simulator can match aggregate summaries despite weak local predictions. Both are evaluated and scored against data of held-out sessions.

\paragraph{Model-Discovery Operators.}\label{sec:operators}
We study autonomous LLM coding agents as \emph{model-discovery operators}. An agent works on its own inside an isolated workspace. Given a task description and the training data, it explores the data,
fits candidate models, and submits a single runnable script, taking on the role a human
modeler usually plays. A fixed \emph{evaluation pipeline} (an \emph{evaluation harness},
in machine-learning terminology) then scores that script on held-out data and records
its cost and process. One key objective is to investigate how the performance, cost, and qualitative behavior of the operator change with the agent and with the reasoning effort we request.

We evaluate two operators: \texttt{codex} (OpenAI Codex CLI v0.125, model GPT-5.5;
\citealp{openaiCodex}) and \texttt{claude-code} (Claude Code, Claude Opus~4.7;
\citealp{anthropicClaude}). Each agent has a native reasoning-effort control, which we
use at three settings (for \texttt{codex}, low, medium, and xhigh; for
\texttt{claude-code}, medium, high, and max). These settings do not mean the same thing
across providers, and within one provider they are ordinal, not numeric. We therefore
treat effort as an ordered within-agent ladder with three levels, which we label low, default, and max in an increasing order. The coding of these effort levels is given in Section~\ref{sec:analysis}.

\subsection{Design Factors and Levels}
\label{sec:design}
We consider a full factorial design over six factors (Table~\ref{tab:factors}) for experiments on the model-discovery operators.
Three are \emph{stratification factors} that set the context in which an effort effect is read: agent $A$, task $T$, and primary metric $M$, each with two levels. The other three
are \emph{design factors} that vary within that context: reasoning effort $E$ and discovery
fold $F$, each with three levels, and discovery regime $R$ with two levels. Thus there are $2\times2\times2\times3\times3\times2 = 144$ runs in total, each a single agent execution under one fully specified configuration. The design has no replicates. Each configuration is run once. Replicating runs at a larger scale, which would enhance uncertainty quantification and give more stable inference, is left to future work.

\begin{table}[htbp]
\centering\small
\caption{The six crossed factors, grouped by role. The three stratification factors
($A,T,M$) index the eight analysis strata, within which the effect of effort is
estimated. The three design factors ($E,F,R$) vary inside a stratum. Their full crossing
gives $144$ runs.}
\label{tab:factors}
\begin{tabular}{llll}
\toprule
Role & Notation & Factor & Levels \\
\midrule
\multirow{3}{*}{Stratification}
 & $A$ & Agent            & \texttt{codex}, \texttt{claude-code} \\
 & $T$ & Task             & \texttt{timestep} (Task~1), \texttt{abm} (Task~2) \\
 & $M$ & Primary metric   & Task~1: wAUC, MRI-RO;\ \ Task~2: $\mathrm{KL}_6$, $D_{\mathrm{LD}}$ \\
\midrule
\multirow{3}{*}{Design}
 & $E$ & Reasoning effort & ordered ladder: low, default, max \\
 & $F$ & Discovery fold   & $\mathcal{F}_1,\mathcal{F}_2,\mathcal{F}_3$ \\
 & $R$ & Discovery regime & full, partial \\
\bottomrule
\end{tabular}
\end{table}

For example, one run uses the \texttt{codex} agent on the \texttt{timestep} task with
wAUC as its primary metric, at default reasoning effort, holding out fold
$\mathcal{F}_1$, in the partial discovery regime. The discovery regime $R$ controls how much training data the agent sees. Fix a held-out
fold, say $\mathcal{F}_1$. The full regime gives the agent the other two folds,
$\mathcal{F}_2\cup\mathcal{F}_3$, for discovery. The partial regime gives only one of
those two folds, chosen in cyclic order ($\mathcal{F}_2$ when $\mathcal{F}_1$ is held
out), which halves the visible training data. In both regimes the held-out fold
$\mathcal{F}_1$ is never shown to the agent and is used only for scoring. A full run and
a partial run that share the same task, metric, effort, and fold are therefore matched. They differ only in how much training data the agent saw, with one exception for the
codex Task-1 runs noted in Appendix~\ref{sec:impl}. This matching is what makes the
full-versus-partial comparison in Section~\ref{sec:res-eda} meaningful for the remaining
strata.

It is seen that effort is only interpretable within an agent, and the metrics live on per-task scales,
so $A$, $T$, and $M$ are not effects to average over but stratification factors that fix the context for reading the effort effect. The natural unit of analysis is therefore a
single \emph{stratum} rather than the full $144$-run array. We call each task--metric combination a \emph{condition} and abbreviate the four conditions
as \texttt{t-wAUC}, \texttt{t-MRI}, \texttt{a-KL}, and \texttt{a-DLD}
(these four metrics are defined in Section~\ref{sec:metrics}), with the prefix
\texttt{t-} for the \texttt{timestep} task and \texttt{a-} for the \texttt{abm} task. (Throughout, we reserve the word \emph{regime} for the
full-versus-partial discovery regime.) A stratum is one condition for one agent. The three design factors vary within it, so it holds $18$ runs ($3$ effort levels $\times\,3$
folds $\times\,2$ regimes). The whole experimental design has eight strata. Note that there are four Task-2 runs that fail the simulation-validity check of Appendix~\ref{sec:impl} and are dropped, leaving $140$ valid runs. All these four runs are under the \texttt{a-DLD} condition.

\subsection{Model-Discovery Targets and Experimental Outputs}
\label{sec:metrics}

\paragraph{Model-Discovery Target Metrics for Agents.}
In each run, the agent is instructed in its prompt to optimize one designated metric, denoted its \emph{target metric} (the primary metric $M$ of Table~\ref{tab:factors}). Note that our evaluation pipeline still computes both metrics for the run, not only the target metric. It can help quantify how the choice of target shifts performance on the metrics a run was not optimizing. The
two Task-1 metrics follow the next-action prediction evaluation in our earlier work on these experiments \citep{he2025model}. The two Task-2 metrics follow the
simulation-adequacy criteria in the group-anagram modeling literature \citep{cedenomieles2019anagram,ren2018generative}. Each metric is computed from the single run of one configuration. For notation convenience, we drop the configuration index and keep only the fold superscript $(k)$.

\medskip\noindent\textbf{Weighted AUC (wAUC, Task~1).}
Task~1 is heavily imbalanced because idle actions dominate each session in the data. Thus it is useful to evaluate a predictive model by how well it ranks the true action above the others, class by class, rather than by raw accuracy. For each held-out observation $r=(v,t)$ let $y_{r}=a_{v,t+1}$ be
the true action and $\widehat p_{rj}=\widehat p(a_{v,t+1}=j\mid h_{vt})$ the predicted
probability of class $j$. For fold $k$, let $n_{j}^{(k)}$ be the number of held-out observations
with $y_{r}=j$ and $n^{(k)}=\sum_{j=1}^{4}n_{j}^{(k)}$. The one-vs-rest AUC for class $j$
is
\begin{equation}
\mathrm{AUC}_{j}^{(k)}
=\Pr\!\left(\widehat p_{rj}>\widehat p_{r'j}\,\bigm|\,y_{r}=j,\ y_{r'}\neq j\right),
\label{eq:auc}
\end{equation}
where $r$ and $r'$ are drawn independently and uniformly from the held-out observations. The weighted AUC averages these
class AUCs, weighting each by its prevalence \citep{handtill2001simple},
\begin{equation}
\mathrm{wAUC}^{(k)}
=\sum_{j=1}^{4}\frac{n_{j}^{(k)}}{n^{(k)}}\,\mathrm{AUC}_{j}^{(k)} .
\label{eq:wauc}
\end{equation}
Higher values mean better discrimination across all four classes at once.

\medskip\noindent\textbf{Mean Relative Improvement on Rare Observations (MRI-RO, Task~1).}
Because idle dominates, the behaviorally interesting actions, reply, request, and form,
are rare, and a model can score well on wAUC while still putting little probability on
them. MRI-RO targets these rare actions directly: on rare-event observations it measures how much
more probability the model puts on the true action than a simple frequency baseline does.
With rare class set $\mathcal{R}=\{2,3,4\}$ (\emph{reply}, \emph{request}, \emph{form}),
the fold-$k$ baseline is the class frequency over the observations of the discovery
split,
\begin{equation}
\widehat e_{j}^{(k)}
=\frac{\sum_{r\in\mathcal{D}_{\mathrm{disc}}^{(k)}}\mathbf{1}\{y_{r}=j\}}
{|\mathcal{D}_{\mathrm{disc}}^{(k)}|},
\qquad j=1,\ldots,4 .
\label{eq:mri_baseline}
\end{equation}
On the rare-event held-out observations
$\mathcal{D}_{\mathrm{rare}}^{(k)}=\{r\in\mathcal{D}_{\mathrm{eval}}^{(k)}\colon y_{r}\in\mathcal{R}\}$,
define the per-observation relative improvement
$\Delta_{r}=(\widehat p_{r,y_{r}}-\widehat e_{y_{r}}^{(k)})/\widehat e_{y_{r}}^{(k)}$
and average,
\begin{equation}
\mathrm{MRI}^{(k)}=\frac{1}{|\mathcal{D}_{\mathrm{rare}}^{(k)}|}
\sum_{r\in\mathcal{D}_{\mathrm{rare}}^{(k)}}\Delta_{r}.
\label{eq:mri}
\end{equation}
Zero means the model matches the baseline on true rare events, and positive values mean
improvement. MRI-RO is a task-specific rare-observation measure motivated by the earlier
anagram evaluation \citep{he2025model}.

\medskip\noindent\textbf{KL Divergence on Player-Level Summaries
($\mathrm{KL}_{6}$, Task~2).}
In earlier group-anagram studies \citep{ren2018generative}, the KL divergence is used to compare the simulated and experimental player-level distributions component by component. Here we follow that approach and add one more temporal component, the mean inter-event time. That is, for player $v$, we record
\begin{equation}
\mathbf{z}_{v}=(z_{v1},\,z_{v2},\,z_{v3},\,z_{v4},\,z_{v5},\,z_{v6}),
\label{eq:task2_summary_vector}
\end{equation}
where $z_{v1}$ and $z_{v2}$ are replies received and sent, $z_{v3}$ and $z_{v4}$ requests
received and sent, $z_{v5}$ the number of words formed, and $z_{v6}$ the mean inter-event
time between consecutive non-idle actions (set to $t_{max}+1$ with $t_{max}=300$ when the player has
fewer than two non-idle actions). For each component $m\in\{1,\ldots,6\}$ the evaluation
pipeline fixes a common bin set $\mathcal{B}_{m}$ from the discovery split and applies it
unchanged to both real and simulated values. Let $C_{mb}^{(k)}$ be the real bin counts
over the held-out players and $\widehat C_{mb}^{(k)}$ the simulated bin counts pooled across
seeds. With additive smoothing $\epsilon>0$ identical across runs, the smoothed
probabilities are
\begin{equation}
P_{mb}^{(k)}=\frac{C_{mb}^{(k)}+\epsilon}
{\sum_{b'}C_{mb'}^{(k)}+\epsilon|\mathcal{B}_{m}|},
\qquad
Q_{mb}^{(k)}=\frac{\widehat C_{mb}^{(k)}+\epsilon}
{\sum_{b'}\widehat C_{mb'}^{(k)}+\epsilon|\mathcal{B}_{m}|},
\label{eq:kl_smoothed_distributions}
\end{equation}
the component divergence is
$\mathrm{KL}_{m}^{(k)}=\sum_{b\in\mathcal{B}_{m}}P_{mb}^{(k)}\log\bigl(P_{mb}^{(k)}/Q_{mb}^{(k)}\bigr)$,
and the Task~2 metric averages over the six components,
\begin{equation}
\mathrm{KL}_{6}^{(k)}=\frac{1}{6}\sum_{m=1}^{6}\mathrm{KL}_{m}^{(k)} .
\label{eq:kl6}
\end{equation}
Lower values mean the simulator's player-level summaries match the held-out sessions more
closely.

\medskip\noindent\textbf{Levenshtein-Distance Distribution Distance ($D_{\mathrm{LD}}$, Task~2).}
The second Task~2 metric is to quantify whether simulated players reproduce the local edit-distance structure of human word formation. Mechanistic modeling of these games in \citet{cedenomieles2019anagram} found that human players tend to form each next word
close in Levenshtein distance to the previous one, in line with cognitive load theory \citep{sweller1988cognitive}. We use this signature as a quality criterion. For player
$v$ with time-ordered formed words $w_{v,1},\ldots,w_{v,M_{v}}$, let
$\ell_{v,\tau}=d_{\mathrm{Lev}}(w_{v,\tau},w_{v,\tau+1})$, $\tau=1,\ldots,M_{v}-1$, be the
consecutive Levenshtein distances. Pooling these over the held-out players gives an empirical distribution with CDF $F^{(k)}$. The simulated trajectories give the analogous CDF $\widehat F^{(k)}$. The metric is the one-dimensional Wasserstein-1 distance between them,
\begin{equation}
D_{\mathrm{LD}}^{(k)}=\int_{0}^{\infty}
\bigl|F^{(k)}(z)-\widehat F^{(k)}(z)\bigr|\,dz ,
\label{eq:ld_distance}
\end{equation}
computed in $O(N\log N)$ time, with $N$ the number of pooled distances, by sorting both
samples \citep{flamary2021pot}. Lower values mean the edit-distance structure is
reproduced more closely.

\medskip
To make the four metrics comparable, we map each raw metric to a
direction-normalized score, flipping the sign of the divergence metrics
$\mathrm{KL}_{6}$ and $D_{\mathrm{LD}}$ so that higher always means better. Comparability across tasks then comes from standardizing each response coordinate within its stratum.

\paragraph{Evaluation Protocol and Outcomes.}
Every run is executed in isolation, so that the agent works in its own workspace with a no-leakage rule for held-out data. For every experimental run, the pipeline also records monetary cost, wall-clock time, and process traces, including how many candidate models and features the agent registered, how many
modeling decisions it logged, and how long its submitted script was. Each agent
additionally reports a \emph{self-assessed score} for its submission, its own estimate of
how well the model performs, which we use in the self-assessment analysis of
Section~\ref{sec:res-eda}. Full implementation details are in Appendix~\ref{sec:impl}.

From each experimental run, we collect eight outcome responses $\by\in\R^{8}$ as shown in Table~\ref{tab:outcomes}. The two \emph{performance} outcomes are the
direction-normalized scores on the run's primary and non-primary metrics. The two
\emph{cost} outcomes are total monetary cost (US dollars) and wall-clock time
(seconds), each on the $\log_{10}$ scale. The four \emph{process} outcomes are the numbers of registered models, registered
features, and logged decisions, and the submitted-script length, each on the
$\log_{10}(1+\cdot)$ scale.

\begin{table}[htbp]
\centering\small
\caption{The eight response coordinates of $\by\in\R^{8}$, in three groups, with the
transform applied to each. Each coordinate is standardized to zero mean and unit variance
within its stratum before the analysis of Section~\ref{sec:analysis}.}
\label{tab:outcomes}
\begin{tabular}{lll}
\toprule
Type & Response & Transformation \\
\midrule
Performance & $y^{(1)}$: primary-metric score        & direction-normalized \\
Performance & $y^{(2)}$: non-primary-metric score    & direction-normalized \\
Cost        & $y^{(3)}$: monetary cost               & $\log_{10}$ \\
Cost        & $y^{(4)}$: wall-clock time             & $\log_{10}$ \\
Process     & $y^{(5)}$: registered candidate models & $\log_{10}(1+\cdot)$ \\
Process     & $y^{(6)}$: registered features         & $\log_{10}(1+\cdot)$ \\
Process     & $y^{(7)}$: logged modeling decisions   & $\log_{10}(1+\cdot)$ \\
Process     & $y^{(8)}$: submitted-script length     & $\log_{10}(1+\cdot)$ \\
\bottomrule
\end{tabular}
\end{table}
\section{Analysis Methods}
\label{sec:analysis}

We analyze each of the eight strata separately. A stratum is one condition for one agent. The main analyses compare an agent against itself across effort levels rather than placing the two agents on a shared effort scale. A few descriptive and univariate analyses pool the agents at the task--metric level and include an agent indicator. These pooled analyses summarize cross-agent differences in the shape of the effort response and in operating cost. This section details model
estimation and inference, and the proposed canonical decomposition accommodating the
direction of the effort effect.

\subsection{Model Estimation and Inference}
\label{sec:ana-model}

For run $i$, the eight standardized responses defined in Section~\ref{sec:metrics} form the
vector $\by_i\in\R^{8}$. Note that reasoning effort is an ordered factor with three levels.
We code this factor by the standard orthogonal polynomial contrasts, with the linear
contrast $x^{\mathrm{L}}_i$ taking $-1$, $0$, and $+1$ for the low, default, and max levels,
and the quadratic contrast $x^{\mathrm{Q}}_i$ taking $+1$, $-2$, and $+1$
\citep{wu2011experiments}. We consider the discovery fold and regime as blocking terms. The
regime indicator $R_i$ equals $1$ for partial-regime runs and $0$ otherwise, and
$F^{(2)}_i$ and $F^{(3)}_i$ are indicators for folds $2$ and $3$, with fold $1$ as the
reference. The design row for run $i$ is
$\mathbf{x}_i=(1,\,x^{\mathrm{L}}_i,\,x^{\mathrm{Q}}_i,\,R_i,\,F^{(2)}_i,\,F^{(3)}_i)^{\top}\in\R^{q}$,
$q=6$, and each response follows the linear model
\begin{equation}
\by_i=\bB^{\top}\mathbf{x}_i+\boldsymbol{\varepsilon}_i,\qquad
\boldsymbol{\varepsilon}_i \ \text{i.i.d. } \Norm(\bze,\bSig),
\label{eq:mlm}
\end{equation}
where $\bB\in\R^{q\times 8}$ is the coefficient matrix and $\bSig$ is the error covariance
across the eight responses. Stacking the $n$ runs of a stratum row-wise into
$\bY\in\R^{n\times 8}$ and $\bX\in\R^{n\times q}$ writes \eqref{eq:mlm} in matrix form
$\bY=\bX\bB+\bER$, with rows of $\bER$ i.i.d.\ $\Norm(\bze,\bSig)$ \citep{anderson2003}.
Table~\ref{tab:design} lists the six design columns.

\begin{table}[htbp]
\centering\small
\caption{The six columns of the design matrix $\bX\in\R^{n\times 6}$ in
model~\eqref{eq:mlm}. These are also the rows of $\bB$. The two effort contrasts carry the
effect of interest, and the fold and regime terms are blocking factors.}
\label{tab:design}
\begin{tabular}{lll}
\toprule
Term & Role & Meaning \\
\midrule
$1$                 & intercept         & stratum baseline (mean) \\
$x^{\mathrm{L}}$    & effort, linear    & trend across the levels, coded $-1,0,+1$ \\
$x^{\mathrm{Q}}$    & effort, quadratic & curvature across the levels, coded $+1,-2,+1$ \\
$R$                 & blocking          & partial-regime indicator, $1$ if partial and $0$ if full \\
$F^{(2)},F^{(3)}$   & blocking          & fold indicators, with fold $1$ as the reference \\
\bottomrule
\end{tabular}
\end{table}

In addition to the per-stratum analysis, we adopt a simple pooled model for investigation
of the primary metric on its own. For each of the four conditions we regress the
direction-normalized primary score $y^{\mathrm{(1)}}_i$ on the two effort contrasts, an
agent indicator $A_i$ for the \texttt{claude-code} agent (\texttt{codex} as the reference),
the regime indicator $R_i$, and fold effects $\phi_{f(i)}$. The simple pooled model can be
expressed as
\begin{equation}
y^{\mathrm{(1)}}_i = \beta_0 + \beta^{\mathrm L}\,x^{\mathrm{L}}_i
  + \beta^{\mathrm Q}\,x^{\mathrm{Q}}_i
  + \gamma\,A_i + \delta\,R_i + \phi_{f(i)} + \varepsilon_i .
\label{eq:univ}
\end{equation}
Unlike the per-stratum multivariate model \eqref{eq:mlm}, this model pools the two agents through
$A_i$, and we report its fit in Section~\ref{sec:res-manova}. Notice that we fit this model separately within each condition rather than including two stratification factors $T$ and $M$, because the primary scores are defined on different task--metric scales.

It is easy to obtain the ordinary least squares estimator $\hat\bB$. Under the normal
assumption in \eqref{eq:mlm}, one can also easily get the cross covariance for $\hat\bB$
\citep{anderson2003}. Let us denote $\hat\bbeta_{l}\in\R^{8}$ to be the row of $\hat\bB$
corresponding to the linear-effort effect (for simplicity, it's written as $\hat\bbeta$ in
Section~\ref{sec:results}). Similarly, we can denote $\hat\bB_{E}$ for the two rows of $\hat\bB$
corresponding to the linear and quadratic effort effects. The distributions of
$\hat\bbeta_{l}$ and $\hat\bB_{E}$ can be easily obtained.

To separate the variation that effort explains from the rest, let
$\bH_{\mathrm{f}}=\bX(\bX^{\top}\bX)^{-1}\bX^{\top}$ be the projection onto the full design
and $\bH_{\mathrm{r}}$ the matching projection for the reduced design that drops the two
effort contrasts. The error and effort (hypothesis) sums of squares and cross-products are
\begin{equation}
\SE=\bY^{\top}(\bI-\bH_{\mathrm{f}})\bY,\qquad
\SH=\bY^{\top}(\bH_{\mathrm{f}}-\bH_{\mathrm{r}})\bY .
\end{equation}
$\SE$ is the within-stratum variation that effort does not explain, and $\SH$ is the variation the two effort
contrasts explain.

We are interested in whether reasoning effort moves the response, and in which direction.
Let $\bB_{E}$ denote the two rows of $\bB$ that hold the linear and quadratic effort
effects. We then formulate the hypothesis test as
\begin{equation*}
H_0:\ \bB_{E}=\bze \qquad\text{versus}\qquad H_1:\ \bB_{E}\neq\bze .
\end{equation*}
To examine whether effort moves the response in any direction, we adopt Pillai's trace
\citep{pillai1955} as the test statistic,
\begin{equation}
V=\tr\!\big[\SH(\SH+\SE)^{-1}\big]=\sum_{k}\frac{\lambda_k}{1+\lambda_k},
\label{eq:pillai}
\end{equation}
where $\lambda_1\ge\lambda_2\ge 0$ are the nonzero roots of the generalized eigenproblem
\begin{equation}
\SH\,\ba=\lambda\,\SE\,\ba ,
\label{eq:geig}
\end{equation}
and the $p$-value comes from the usual $F$ approximation.

Moreover, it is of great interest to examine a particular direction for the contrast of
performance and cost responses. Specifically, we consider the utility direction as
\begin{equation}
\bu=\big(\tfrac12,\tfrac12,-\tfrac12,-\tfrac12,0,0,0,0\big)^{\top},
\end{equation}
which emphasizes the two performance coordinates and penalizes the two cost coordinates.
The equal weights here set the tradeoff between performance and cost. Projecting each run
onto $\bu$ gives a scalar utility score $U_i=\bu^{\top}\by_i$. Regressing $U$ on the design factors is equivalent to a univariate regression of $U$
\citep{rencher2012}. Its linear-effort slope is $\bu^{\top}\hat\bbeta_l$, with estimated
variance $c_{\ell\ell}\,\bu^{\top}\hat\bSig\,\bu$, where
$c_{\ell\ell}$ is the $x^{\mathrm L}$ diagonal entry of $(\bX^{\top}\bX)^{-1}$.
We test this slope with a $t$ statistic. A negative slope,
$\bu^{\top}\hat\bbeta_l<0$, means more effort lowers utility along the
pre-specified utility direction. Because $\bu$ is fixed in advance, this directed
test focuses on the utility direction rather than testing all response directions
simultaneously, and it has more power when the effect concentrates along $\bu$
\citep{johnson2007}. Across strata, we apply Holm's multiple-testing correction to the eight per-stratum utility-test $p$-values \citep{holm1979}. As a supplementary analysis, we also fit a pooled model with agent
as a covariate.

\subsection{Utility-Aligned Canonical Decomposition}
\label{sec:ana-uacd}

\begin{figure}[!htb]
\centering
\includegraphics[width=.82\textwidth]{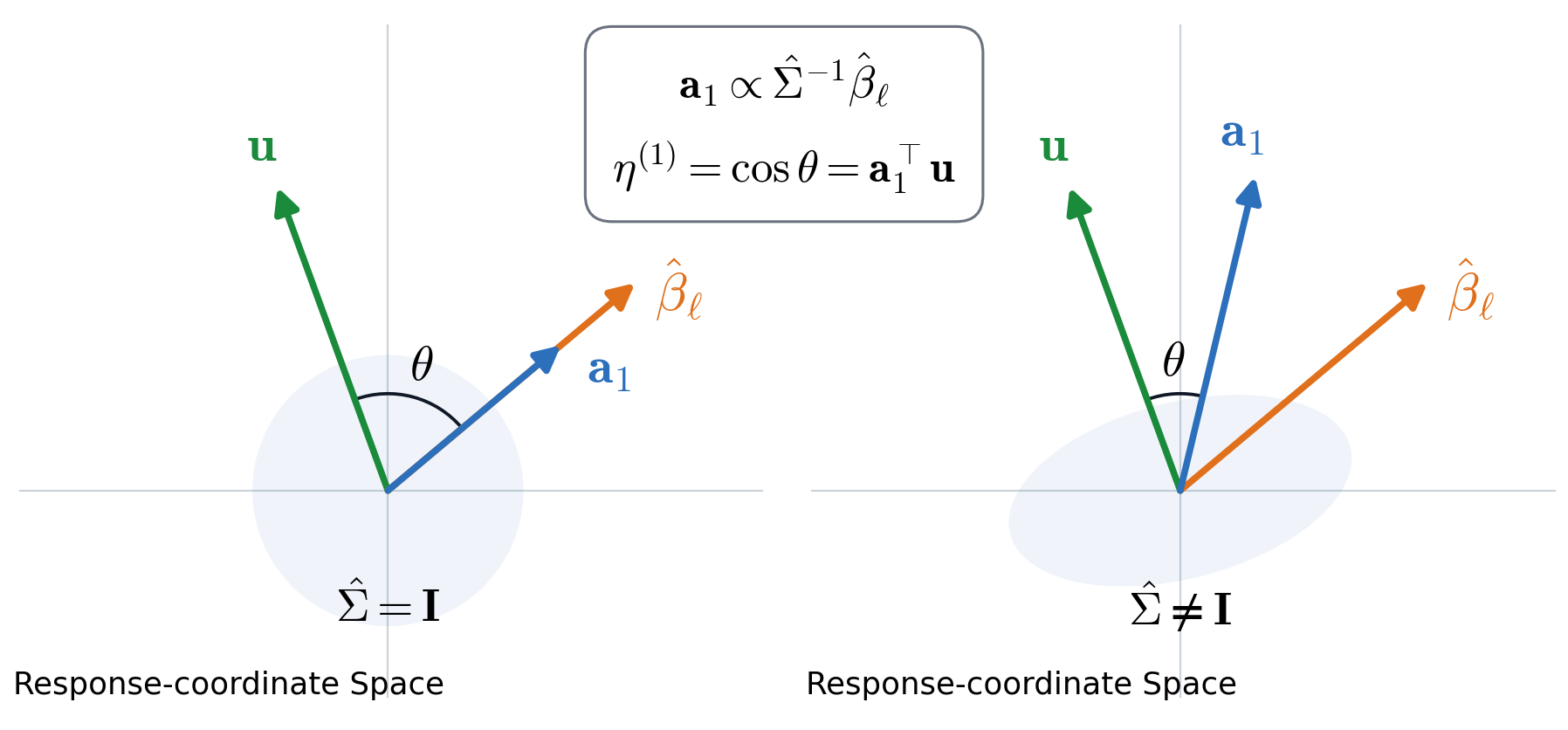}
\caption{Simplified rank-one illustration of UACD. Left panel shows the trivial case where $\hat{\boldsymbol\Sigma}=I$, while the right panel shows the non-trivial case where $\hat{\boldsymbol\Sigma}\neq I$.}
\label{fig:uacd-concept}
\end{figure}

The omnibus test using the test statistics in \eqref{eq:pillai} says whether effort moves the response. The directed
test says whether it moves toward utility. To describe \emph{which} direction effort moves,
we will investigate the leading canonical direction of the effect and compare it with $\bu$.
For the classical canonical decomposition of a MANOVA effect, the eigenvectors of
\eqref{eq:geig} are the \emph{canonical directions} $\ba_1,\ba_2$ \citep{anderson2003}. The
leading one solves
\begin{equation}
\ba_1=\arg\max_{\ba}\ \frac{\ba^{\top}\SH\,\ba}{\ba^{\top}\SE\,\ba} .
\end{equation}
This direction accounts for the effort-explained variation after adjustment by the
residual covariance $\SE^{-1}$, so noisy or redundant response directions receive less
weight. It can be interpreted as the single response composite that effort moves most,
relative to residual variation. We set $\lVert\ba_1\rVert=1$. Since the sign of a canonical direction is arbitrary,
we orient $\ba_1$ so that $\ba_1^{\top}\hat\bbeta_l>0$, where $\hat\bbeta_l$ is the
linear-effort coefficient vector. With this convention, the sign of
$\eta^{(1)}=\ba_1^{\top}\bu$ indicates whether the direction associated with increasing
effort is aligned with the utility direction. We consider two scalars to quantify the concentration and utility alignment of the effort effect,
\begin{equation}
\pi=\frac{\lambda_1}{\lambda_1+\lambda_2}\in[\tfrac12,1],\qquad
\eta^{(1)}=\ba_1^{\top}\bu\in[-1,1].
\end{equation}
Here $\lambda_1\ge\lambda_2$ are the two nonzero roots of \eqref{eq:geig}. The concentration
$\pi$ measures how close to one-dimensional the effort effect is. Thus a value near $1$
means effort acts along a single direction. Note that the effect of two effort contrasts
spans at most two canonical directions. Here we focus on the alignment of the leading
direction alone, since the second direction often plays a small role, as shown in
Section~\ref{sec:res-manova}. The alignment $\eta^{(1)}$ is the cosine of the angle between $\mathbf a_1$ and $\mathbf u$. To clarify its interpretation, consider the rank-one special case in which only the linear effort effect is tested (Figure~\ref{fig:uacd-concept}). In that case, $\mathbf a_1 \propto \hat{\boldsymbol\Sigma}^{-1}\hat{\boldsymbol\beta}_l ,$ where $\hat{\boldsymbol\beta}_l$ is the raw linear-effort effect and $\hat{\boldsymbol\Sigma}$ is the estimated error covariance matrix among response coordinates. If $\hat{\boldsymbol\Sigma}=I$, meaning the residual response coordinates have unit variance and zero covariance, then $\mathbf a_1$ points in the same direction as $\hat{\boldsymbol\beta}_l$. If $\hat{\boldsymbol\Sigma}$ is nontrivial, then $\mathbf a_1$ is a covariance-adjusted version of $\hat{\boldsymbol\beta}_l$. Thus the directed utility test asks whether the raw linear effort trend improves the pre-specified utility score through $\mathbf u^\top\hat{\boldsymbol\beta}_l$, whereas UACD describes the dominant covariance-adjusted direction of the multivariate effort effect and measures its alignment with $\mathbf u$ through $\eta^{(1)}=\mathbf a_1^\top\mathbf u$.

Note that $\ba_1$ is a nonlinear function of $\SH$ and $\SE$ and is estimated from the 18
runs (or 16 runs for the \texttt{a-DLD} condition, where four invalid runs were dropped) of a stratum, so its direction can have large uncertainty. A case-resampling bootstrap
\citep{efron1993} indicates that the leading direction is weakly identified within a single
stratum. We therefore interpret $\eta^{(1)}$ by its sign rather than by a confidence interval. We then
assess whether that sign is consistent across the eight strata with a sign test, against the
null that the leading direction is equally likely to point toward or away from utility. We
treat this as a descriptive pattern that complements, rather than replaces, the directed
utility test. In a short summary, the three quantities play different roles. The directed
utility effect $\bu^{\top}\hat\bbeta_{l}$ is an exact linear functional of $\hat\bbeta_{l}$
and supplies both the effect estimate and its test. Pillai's trace tests for a joint effort
effect in any direction. The triple $(\pi,\ba_1,\eta^{(1)})$ describes the shape of that
effect.
\section{Case Study Results}
\label{sec:results}

We first characterize the experimental results descriptively, including how the agents
specialize, how they spend resources, the effect of full versus partial information, and the accuracy of
their self-assessment (Section~\ref{sec:res-eda}). Section~\ref{sec:res-manova} then reports formal inference on the treatment effect of reasoning effort. We examine this effect on the primary metric alone, through a pre-specified utility contrast, and through an omnibus multivariate test interpreted with the utility-aligned canonical decomposition (UACD). All inferential statements are conditional on the $n=140$ valid runs.

\subsection{Exploratory Analysis}
\label{sec:res-eda}

Before turning to formal inference, we describe the experiment to show how the two agents
behave under the reasoning-effort ladder and to reveal the patterns that the multivariate
analysis of Section~\ref{sec:res-manova} accounts for. The analysis sample is the $140$
simulation-valid runs of the $144$ scheduled. The validity screen, the balance across agents
and conditions, and the four excluded runs are detailed in Appendix~\ref{sec:app-extra}.

\begin{figure}[htbp]\centering
\includegraphics[width=.85\textwidth]{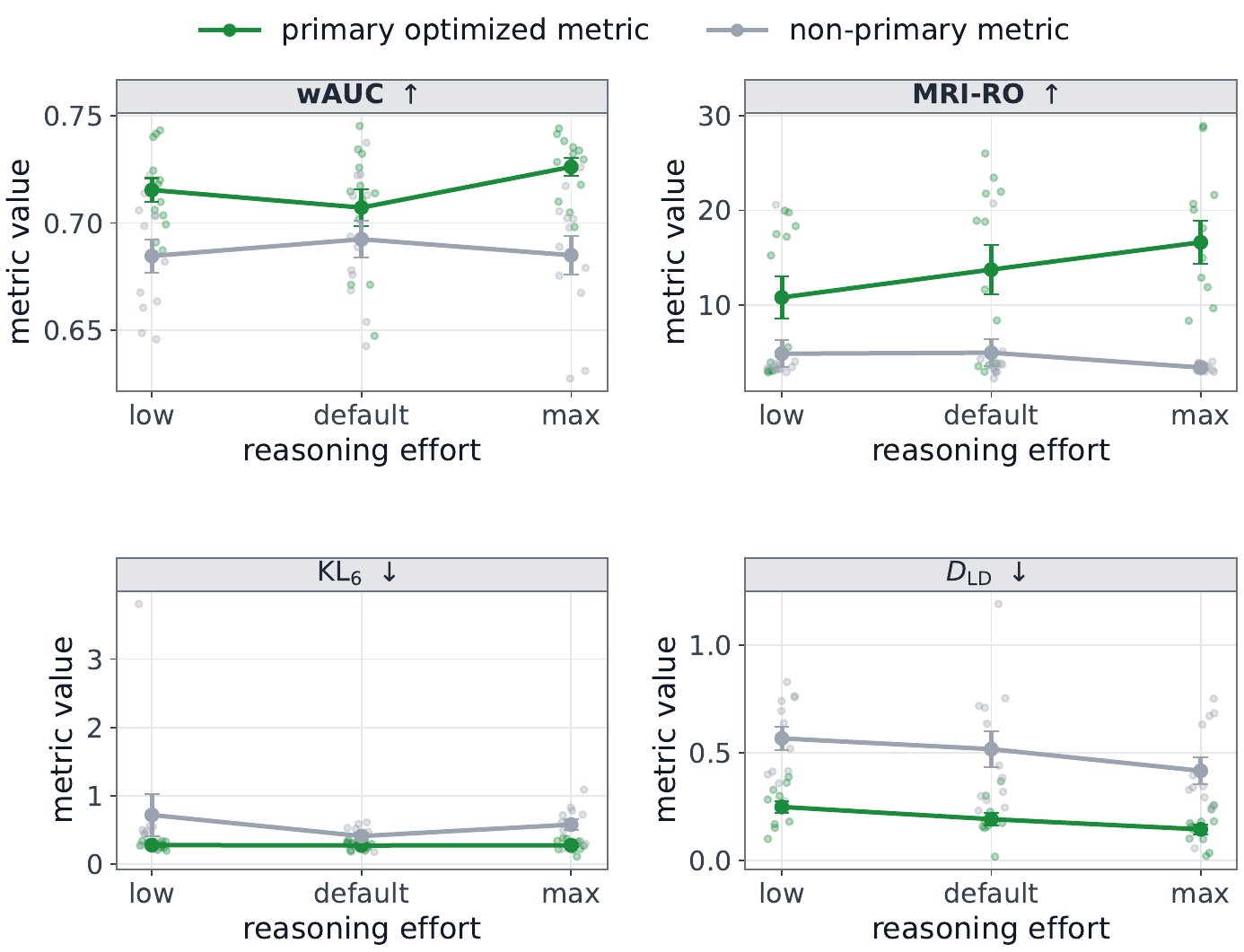}
\caption{Performance by primary metric across the reasoning-effort ladder. Each
metric is split into the runs that optimized it and the runs where it was only recorded.}
\label{fig:target}
\end{figure}

Recall that the agents optimize the metric they are assigned. Figure~\ref{fig:target} plots
each metric across the effort ladder, with group means joined by lines, $\pm 1$
standard-error bars, and individual runs as points. Within each metric the runs that
optimized it are separated from those where it was only recorded. It is clear that the
primary series outperforms the non-primary one in every metric, and the gap is largest for
MRI-RO. The descriptive pattern for reasoning effort is metric-dependent. It
improves descriptively for MRI-RO and for $D_{\mathrm{LD}}$, while weighted AUC and
$\mathrm{KL}_6$ stay flat. Overall, the agents show strong metric specialization, whereas the apparent performance gains from reasoning effort are limited to some metrics.

\begin{figure}[htbp]\centering
\includegraphics[width=.95\textwidth]{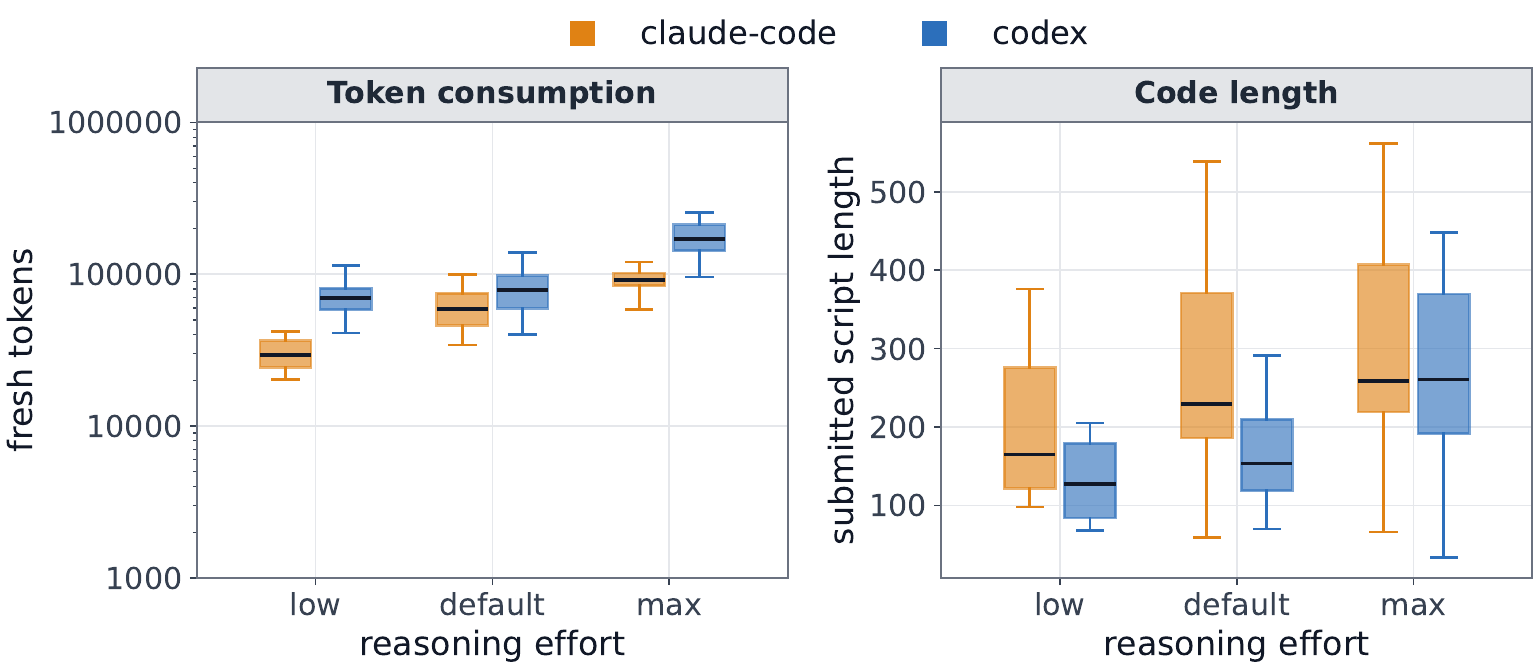}
\caption{Fresh-token use (left) and submitted-code length (right) across the reasoning-effort
ladder, by agent.}
\label{fig:resources}
\end{figure}

The clearest effect of reasoning effort is on resource use. Figure~\ref{fig:resources} shows
box plots of fresh-token totals and submitted-code length at each effort level for the two
agents. Here fresh tokens are the non-cached input plus output tokens; the cached pool
(cache-creation and cache-read tokens) is excluded and tracked separately. One can see that
both quantities grow as effort rises. Median fresh-token use climbs from $29.4$k to $91.6$k
for claude-code and from $69.6$k to $170.5$k for codex between the lowest and highest setting.
Codex uses more fresh tokens at every level yet costs less in dollars (see below), because
most of claude-code's cost comes from cache reads that the fresh-token count omits. Code
length grows in step: claude-code writes the longer script at low effort (median $165$
against $127$ lines), and the two reach about $260$ lines at maximum effort. The number of
candidate models an agent reports considering grows in the same way, which we report in
Appendix~\ref{sec:app-extra}.

\begin{figure}[htbp]\centering
\includegraphics[width=.95\textwidth]{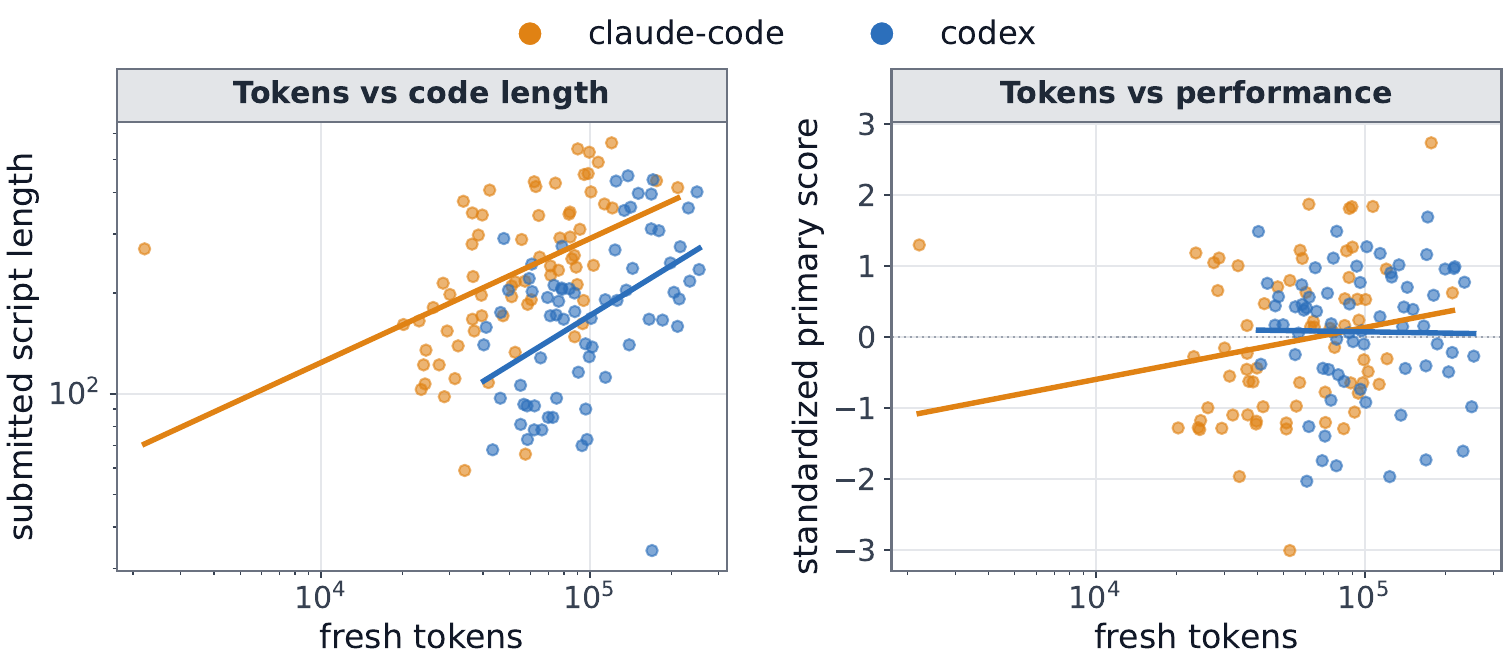}
\caption{Fresh-token use against submitted-code length (left) and against within-condition
standardized performance (right), by agent.}
\label{fig:tokenperf}
\end{figure}

We also find that more tokens reliably produce more code, but not better performance. The
left panel of Figure~\ref{fig:tokenperf} plots submitted-script length against fresh tokens,
with one fitted line per agent, and token use and code length move together. The right panel
places all four conditions on one axis. It standardizes the primary metric within each
condition, orients it so that higher is always better, and plots it against fresh tokens. For
claude-code the token-performance relationship is weakly positive (Spearman $\rho = 0.28$),
and for codex it is essentially flat ($\rho = 0.01$). This anticipates the negative
effort-to-utility slope of Section~\ref{sec:res-manova}.

\begin{figure}[htbp]\centering
\includegraphics[width=.82\textwidth]{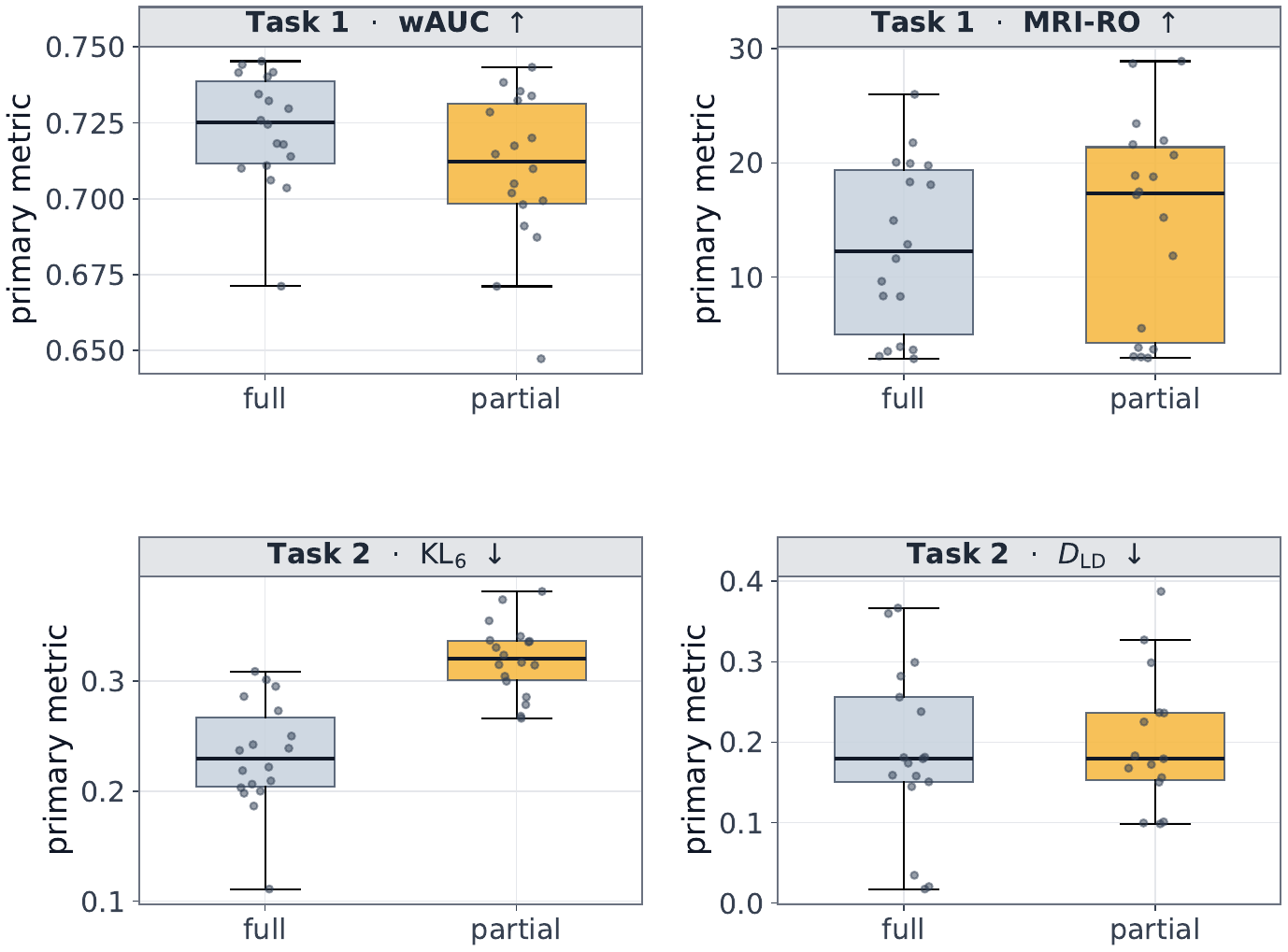}
\caption{Primary performance under full versus partial observation, by task and metric.}
\label{fig:partial}
\end{figure}

The full discovery split shows an advantage over the partial split in some conditions, but not in others. Figure~\ref{fig:partial} places the full- and
partial-information runs side by side for each task and metric as box plots of the raw
primary metric, with individual runs overlaid and axes and direction markers as in
Figure~\ref{fig:target}. The full-information advantage is clear only in \texttt{a-KL}, where the
partial $\mathrm{KL}_6$ values sit well above (worse than) the full ones. The remaining
conditions show overlap, with at most a slight \texttt{t-wAUC} gap. We treat the Task-1 full-versus-partial comparison as descriptive only, because the
prompt-wording caveat of Appendix~\ref{sec:impl} coincides with the codex Task-1 split.

\begin{figure}[htbp]\centering
\includegraphics[width=.75\textwidth]{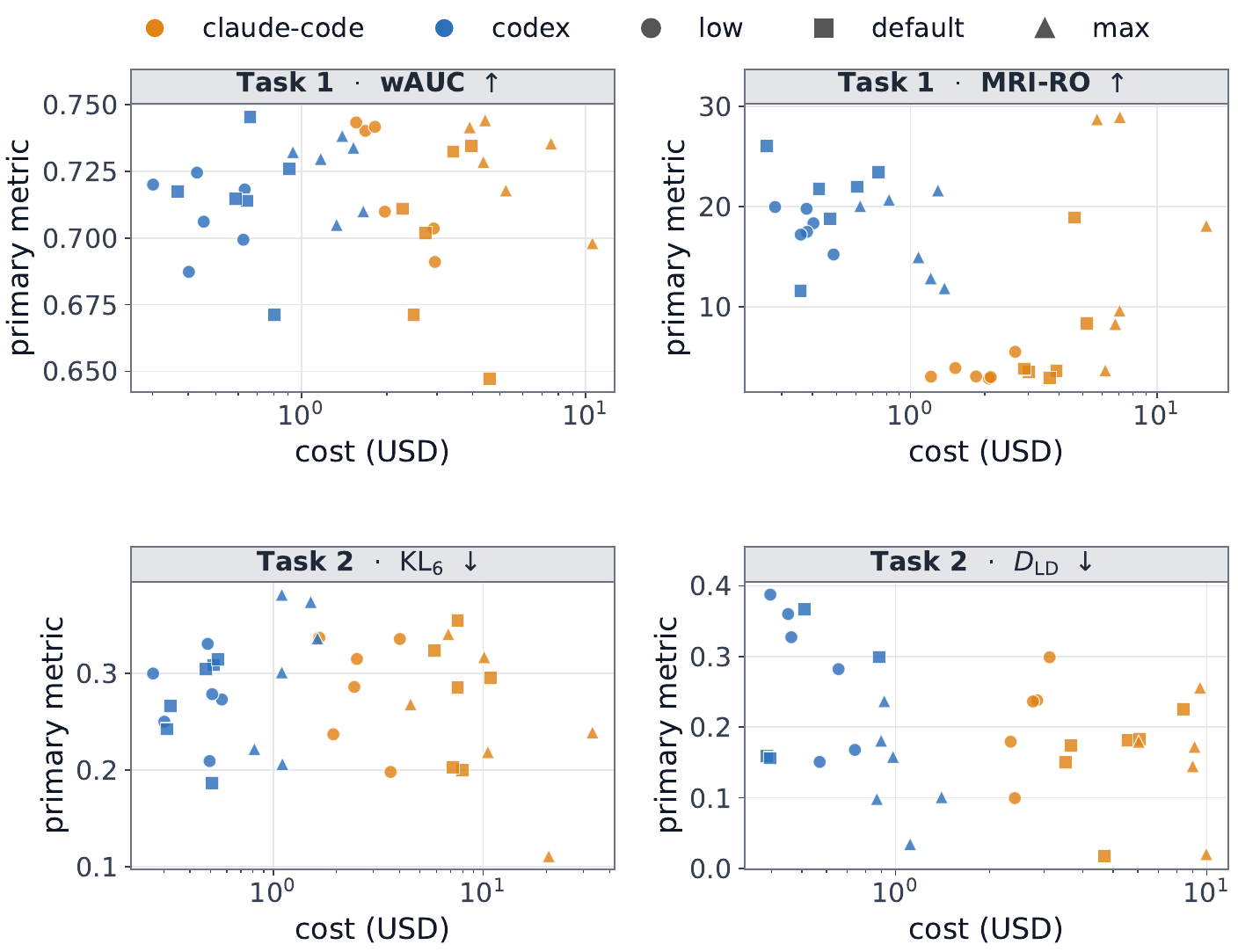}
\caption{Primary metric against cost in US dollars (log axis), by condition. Color for agent and
marker shape for effort.}
\label{fig:cost}
\end{figure}

Note that the two agents occupy different regions of the cost-quality plane, as shown in
Figure~\ref{fig:cost}, which plots the raw primary metric against cost in US dollars. Codex
stays at low dollar cost in every condition, with a median cost of \$$0.60$ against
claude-code's \$$4.19$, roughly seven times higher, while reaching comparable primary-metric
quality. Within an agent, higher cost is not consistently associated with better quality. The
clear exception is claude-code on \texttt{t-MRI}, where higher-cost runs do reach higher MRI-RO
(Spearman $\rho = 0.71$). Because codex dollar costs are imputed rather than metered, this
cross-agent comparison describes deployment economics, not a controlled contrast, though the
ordering holds in all four conditions.

\begin{figure}[htbp]\centering
\includegraphics[width=.75\textwidth]{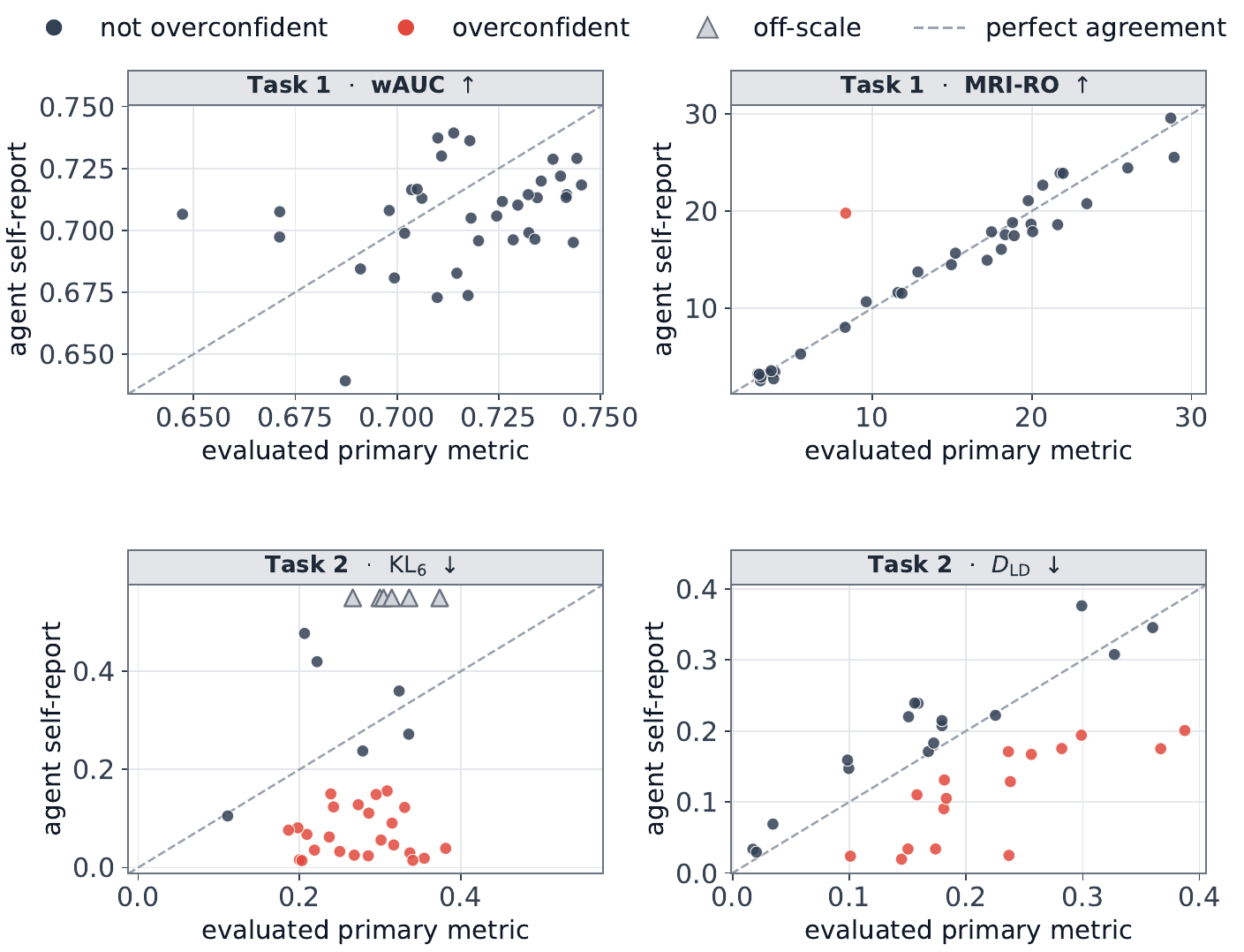}
\caption{Agent self-assessment against the evaluated metric, one panel per condition.}
\label{fig:calib}
\end{figure}

It is found that the agents assess their own work well on the predictive task (Task~1) and poorly on the
generative one (Task~2). Figure~\ref{fig:calib} plots each agent's internal
validation score against the evaluated primary metric assigned by the evaluation pipeline. On
Task~1 the self-reports lie along the diagonal, and only one of the $72$ runs (claude-code on
\texttt{t-MRI}) is flagged overconfident. On Task~2 the self-reports sit on the optimistic side.
Because $\mathrm{KL}_6$ and $D_{\mathrm{LD}}$ are lower-better, the overconfident runs fall
below the diagonal, and they are common ($83\%$ of claude-code's \texttt{a-KL} runs and about half
of codex's), with no sign of improving as effort rises. We also observe that six codex \texttt{a-KL}
runs report internal scores on a scale far from the evaluated metric. These are shown as gray
off-scale markers and reflect the differing internal-validation schemes across runs, so the
plot is best read as a descriptive comparison of self-report against the evaluated metric, not
a formal assessment of self-report accuracy. Self-assessment is reliable for Task~1 but unreliable for Task~2.

The composition of the submitted model families, which is largely agent-specific on Task~1
and spreads across four simulator families on Task~2, is reported in
Appendix~\ref{sec:app-extra}. The four runs left out of the analysis are examined there as
well, as a short case study of how agentic discovery breaks down.

\subsection{Inference on the Treatment Effect}
\label{sec:res-manova}

\paragraph{Univariate primary metric.} We first fit the pooled univariate
model~\eqref{eq:univ} of Section~\ref{sec:ana-model}, regressing the direction-normalized
primary score on the two effort contrasts, the agent indicator $A$, and the blocking
covariates, in each of the four conditions. It is seen that the effort effect on the primary
metric alone is weak (Table~\ref{tab:univ}): the joint effort test reaches uncorrected
significance only in \texttt{a-DLD} ($p$-value $0.029$, which is not significant at
$\alpha/4=0.0125$). The significant effects in the table come from the covariates, not from
effort. The clearest is the large difference between the two agents in \texttt{t-MRI} (the $A$ term,
$p$-value $<0.001$). Effort has little effect on the metric each agent was assigned to
optimize, which motivates the multivariate view, in which effort is allowed to act on cost
and process as well as on performance.

\begin{table}[htbp]
\centering\small
\caption{Univariate primary-metric regression~\eqref{eq:univ} by condition, grouped by task. Here $p_A$ and $p_R$ are the agent ($A$) and partial-regime ($R$) $p$-values. Bonferroni threshold
$\alpha/4=0.0125$.}
\label{tab:univ}
\begin{tabular}{llcccc}
\toprule
 & Condition & $F_{E}$ ($p$-value) & $\beta^{\mathrm L}$ ($p$-value) & $p_{A}$ & $p_{R}$ \\
\midrule
\multirow{2}{*}{Task~1}
 & \texttt{t-wAUC} & $2.84\ (0.075)$ & $+0.005\ (0.187)$ & $0.997$ & $0.054$ \\
 & \texttt{t-MRI}  & $2.40\ (0.108)$ & $+2.92\ (0.037)$  & $<0.001$ & $0.314$ \\
\midrule
\multirow{2}{*}{Task~2}
 & \texttt{a-KL}   & $0.08\ (0.926)$ & $+0.001\ (0.836)$ & $0.288$ & $<0.001$ \\
 & \texttt{a-DLD}  & $4.07\ (0.029)$ & $+0.051\ (0.009)$ & $0.107$ & $0.686$ \\
\bottomrule
\end{tabular}
\end{table}

\begin{figure}[htbp]\centering
\includegraphics[width=\textwidth]{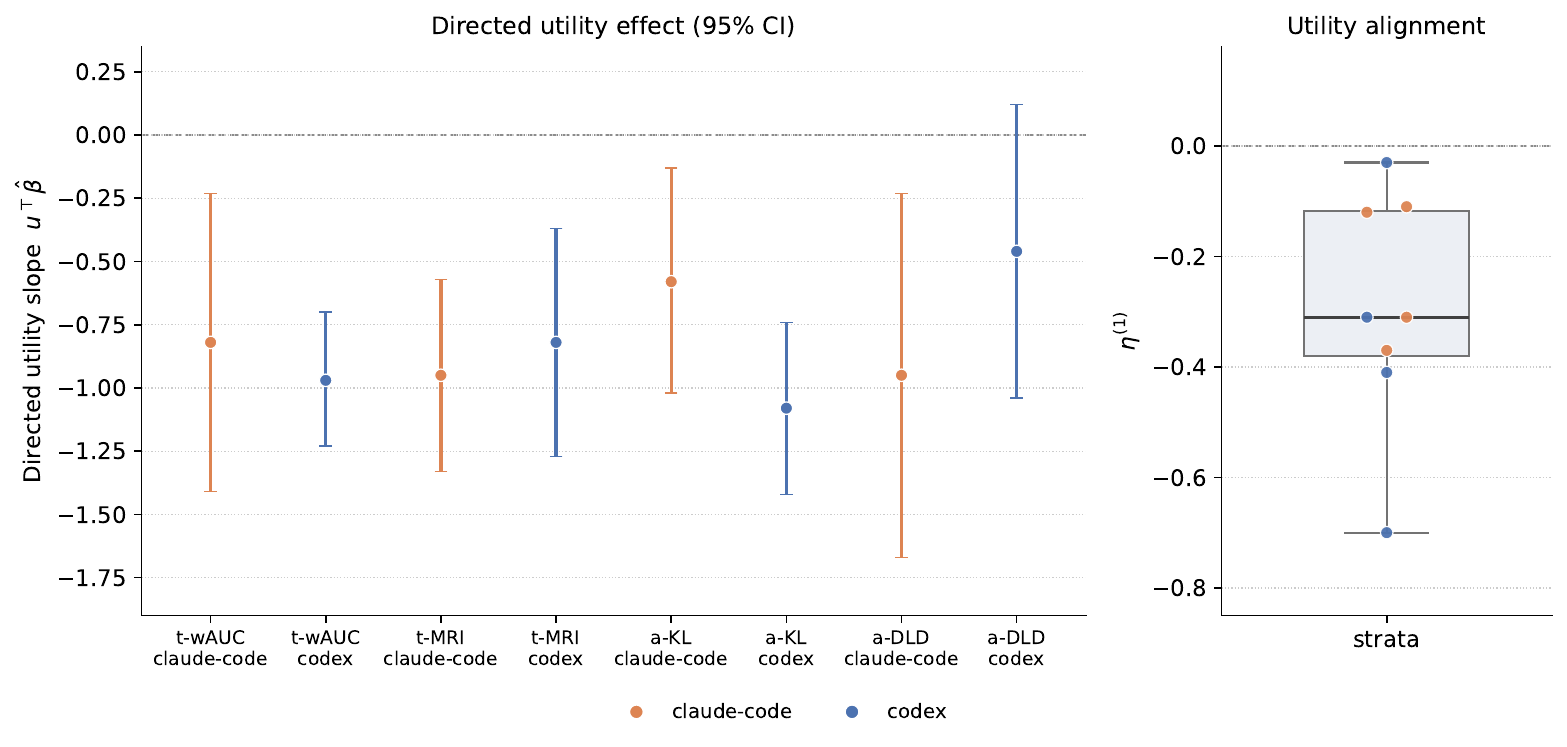}
\caption{Left: directed utility slope $\bu^{\top}\hat\bbeta$ per stratum with its model-based $95\%$ $t$ interval. Right: utility alignment $\eta^{(1)}$ across strata, as a boxplot over the eight point estimates. Color marks the agent.}
\label{fig:eta}
\end{figure}

\paragraph{Directed Utility Effect.} Projecting each run onto the pre-specified utility direction
gives the scalar utility score $U_i=\bu\T\by_i$, and regressing $U$ on the design yields the linear-effort slope $\bu\T\hat\bbeta$, which is the change in utility per unit of effort, and the directed test specified in Section~\ref{sec:analysis}. It is seen that more effort lowers utility in \emph{every} stratum: the slope is
negative in all eight (Table~\ref{tab:directed}; Figure~\ref{fig:eta}, left), and because $\bu\T\hat\bbeta$ is a linear
functional of $\hat\bbeta$, its model-based $t$ interval excludes zero in seven of the eight (all
but codex/\texttt{a-DLD}). The linear-effort slope is significant without correction in seven of eight strata, and
the same seven remain significant after Holm correction across the eight tests. Entering agent ($A$) as a covariate rather than stratifying on it ($n=36$ per condition,
$32$ for \texttt{a-DLD}) leaves the directed effect significant in three of
the four conditions. The exception is \texttt{a-DLD}, where the pooled linear
slope is $-0.27$ with a $p$-value of $0.11$. For those three conditions the negative utility slope
is not merely an artifact of the small per-stratum samples.

\begin{table}[htbp]
\centering\small
\caption{Directed utility effect of effort, per stratum and grouped by task: linear-effort slope $\bu\T\hat\bbeta$ of $U=\bu\T\by$ with its model-based $95\%$ $t$ interval, the slope $p$-value $p_{\mathrm{lin}}$, and its Holm-adjusted value $p_{\mathrm{Holm}}$ across the eight strata.}
\label{tab:directed}
\begin{tabular}{lllccccc}
\toprule
 & Condition & Agent & $n$ & $\bu\T\hat\bbeta$ & $95\%$ CI & $p_{\mathrm{lin}}$ & $p_{\mathrm{Holm}}$ \\
\midrule
\multirow{4}{*}{Task~1}
 & \texttt{t-wAUC} & claude-code & 18 & $-0.82$ & $[-1.41,-0.23]$ & $0.011$  & $0.042$  \\
 & \texttt{t-wAUC} & codex       & 18 & $-0.97$ & $[-1.23,-0.70]$ & $<0.001$ & $<0.001$ \\
 & \texttt{t-MRI}  & claude-code & 18 & $-0.95$ & $[-1.33,-0.57]$ & $<0.001$ & $<0.001$ \\
 & \texttt{t-MRI}  & codex       & 18 & $-0.82$ & $[-1.27,-0.37]$ & $0.002$  & $0.010$  \\
\midrule
\multirow{4}{*}{Task~2}
 & \texttt{a-KL}   & claude-code & 18 & $-0.58$ & $[-1.02,-0.13]$ & $0.015$  & $0.044$  \\
 & \texttt{a-KL}   & codex       & 18 & $-1.08$ & $[-1.42,-0.74]$ & $<0.001$ & $<0.001$ \\
 & \texttt{a-DLD}  & claude-code & 16 & $-0.95$ & $[-1.67,-0.23]$ & $0.015$  & $0.044$  \\
 & \texttt{a-DLD}  & codex       & 16 & $-0.46$ & $[-1.04,+0.12]$ & $0.105$  & $0.105$  \\
\bottomrule
\end{tabular}
\end{table}

\paragraph{Omnibus Test and UACD.}  For examining whether effort moves the response vector in \emph{any} direction, the omnibus Pillai trace appears to be large in every stratum ($V\in[1.31,1.66]$, Table~\ref{tab:uacd}). But note that, with $n\in\{16,18\}$ against $d=8$, its per-stratum power is limited as four of eight strata reach uncorrected $p$-value $<0.05$ and none is significant at $\alpha/8=0.00625$. The omnibus test is less targeted for the utility question. It reaches significance in four of eight strata before correction and in none after correction, whereas the directed test reaches significance in seven strata before and after Holm correction. This is consistent with the expected power advantage of a pre-specified directed contrast when the effect concentrates along $\bu$ \citep{johnson2007}. As a pooled supplementary analysis, the same agent-as-covariate pooling ($n=36$ or $32$ for \texttt{a-DLD} condition) makes the omnibus effect significant in all four conditions. This pooling increases sample size, but it also standardizes the cost coordinate after combining two agents with different cost distributions.

The effort effect is concentrated in one leading canonical direction in every stratum, with that direction carrying $\pi\in[0.84,0.99]$ of the canonical effect. Its leading direction is utility-misaligned throughout, with $\eta^{(1)}<0$ in all eight as shown in Figure~\ref{fig:eta}. The dominant covariance-adjusted direction of the effort effect is aligned more with higher cost and process complexity than with improved performance. We can further interpret $\eta^{(1)}$ by its sign rather than by a per-stratum interval (Section~\ref{sec:ana-uacd}). All eight values of $\eta^{(1)}$ are negative. A sign test gives $p=0.008$ under the simplifying assumption that the eight signs are independent and equally likely to be positive or negative. This assumption is only approximate, because the strata share the same agents, tasks, and folds. We therefore treat this sign pattern as descriptive support for the directed contrast, not as a separate basis for inference. This negative sign pattern also persists when cost is defined differently. Replacing USD with fresh-token count leaves $\eta^{(1)}<0$ in all eight strata, while formal inference about the utility effect relies on the directed contrast of Table~\ref{tab:directed}.

The utility direction $\bu$ represents one pre-specified trade-off between performance and cost. The negative directed slopes in Table~\ref{tab:directed} should therefore be interpreted relative to this choice of utility weights. A systematic sensitivity analysis over alternative utility directions is left to future work.

\begin{table}[htbp]
\centering\small
\caption{Eight-stratum omnibus effort test and shape descriptors, grouped by task: Pillai $V$ ($F$-approximation $p$-value $p_{V}$), leading canonical share $\pi$, and utility alignment $\eta^{(1)}$ (read by sign; Section~\ref{sec:ana-uacd}).}
\label{tab:uacd}
\begin{tabular}{lllccccc}
\toprule
 & Condition & Agent & $n$ & $V$ & $p_{V}$ & $\pi$ & $\eta^{(1)}$ \\
\midrule
\multirow{4}{*}{Task~1}
 & \texttt{t-wAUC} & claude-code & 18 & 1.45 & 0.117 & 0.84 & $-0.37$ \\
 & \texttt{t-wAUC} & codex       & 18 & 1.65 & 0.017 & 0.95 & $-0.70$ \\
 & \texttt{t-MRI}  & claude-code & 18 & 1.58 & 0.040 & 0.88 & $-0.12$ \\
 & \texttt{t-MRI}  & codex       & 18 & 1.60 & 0.029 & 0.94 & $-0.31$ \\
\midrule
\multirow{4}{*}{Task~2}
 & \texttt{a-KL}   & claude-code & 18 & 1.31 & 0.276 & 0.88 & $-0.11$ \\
 & \texttt{a-KL}   & codex       & 18 & 1.66 & 0.015 & 0.99 & $-0.41$ \\
 & \texttt{a-DLD}  & claude-code & 16 & 1.55 & 0.225 & 0.91 & $-0.31$ \\
 & \texttt{a-DLD}  & codex       & 16 & 1.54 & 0.231 & 0.99 & $-0.03$ \\
\bottomrule
\end{tabular}
\end{table}

\FloatBarrier
\section{Discussion and Conclusion}
\label{sec:discussion}

\subsection{Summary of Findings}

The main practical implication is that requested reasoning effort acts more like a
resource-control parameter than a reliable path to better models. Higher effort produces
longer, costlier, and more complex runs. It does not produce a matching gain in model
quality. This pattern is clearly reflected through our multivariate analysis. An analysis of performance as a single response
alone would miss this pattern, because the main changes with effort appear in cost and
process rather than in quality. By treating cost and process as multiple response coordinates, we can
ask whether the effort effect moves in a useful direction. In this experiment, the directed
utility slopes show that it does not.

The proposed design framework also clarifies three operating features of these agents. First, the agents
specialize to the metric they are scored on. Gains on the target metric can come with weaker
performance on the other metric, so the choice of objective is itself a consequential design
decision. Second, more training data helps mainly when the objective rewards matching the held-out distribution of outcomes. For the other objectives, the full and partial discovery
regimes show only small differences. Third, the two agents reach comparable primary-metric
quality, but their estimated dollar costs can differ by about a factor of seven. In this testbed,
quality and cost are therefore not tightly coupled across agents.

Another cautionary finding concerns self-assessment. On the predictive task, the agents'
self-reported scores were close to their evaluated scores. On the generative task, the agents
were often optimistic. All four discarded runs came from the generative distance condition.
Appendix~\ref{sec:app-extra} gives the details. The most extreme discarded run was a
maximum-effort run whose simulator was almost entirely rejected by the evaluation pipeline,
although its internal check reported success. Higher effort did not prevent this failure and
did not make the self-report more cautious. For deployment, an agent's self-report should
not replace an external evaluation check, especially for generative tasks.

\subsection{Limitations and Future Work}

There are several limits in the current work. First, the proposed design has not considered replications,
so the per-stratum inference is based on the factorial structure rather than repeated executions
of the same configuration. This may limit the stability of the uncertainty estimates. Note that replication
at larger scale is costly for agentic AI in our setting. The most expensive single run in our
experiment required more than 70 minutes and more than 33 metered dollars for agent work
alone, and the simulation and validation pipeline adds further time effort. Larger replicated studies
are therefore left to future work. Second, our current work investigates only two agents on one
family of discovery tasks. The proposed framework is therefore demonstrated as a technique, not as a survey of agents in general. Third, the dollar cost for one agent is imputed from token counts at list prices rather
than metered directly. For this reason, the cross-agent cost comparison describes deployment
economics and is not a controlled contrast. Fourth, the leading canonical direction is also
weakly identified within a single stratum. We therefore treat the utility-misalignment result
as descriptive, supported by the consistency of its sign across the eight strata.

There are several directions for future work. One can study a wider set of agents
and discovery domains to test how the pattern between reasoning effort and outcomes will
generalize. Future experiments should also record metered cost, so that price can be analyzed
more directly. Another direction is to study the recorded run traces in more detail. This may
show where the extra reasoning effort goes and why it produces little gain in utility. More
generally, further experiments and analyses can be conducted to investigate agentic operators on their behaviors with respect to performance, quality, cost and process in a joint and comprehensive manner.
\bibliographystyle{asa}
\bibliography{refs}
\appendix
\section{Implementation Details}
\label{sec:impl}

This appendix gives the run mechanics that Section~\ref{sec:metrics} summarizes.

\paragraph{Orchestration.}
An orchestrator runs each configuration in three stages. First, it sets up an isolated
workspace that holds the training data for the run's fold and regime, together with a
run-specific prompt. The prompt states the task, the target metric, and the output
contract the submitted script must meet. Second, it launches the agent at the assigned
reasoning effort and lets it work on its own. The agent explores the data, registers the
candidate models and features it tries, records its modeling decisions, and submits one
runnable script. Third, it scores that script on the held-out fold with the metrics of
Section~\ref{sec:metrics}. The agent never sees the held-out fold during discovery.

\paragraph{No-Leakage Rule (Task~1).}
When the evaluation pipeline scores a Task-1 prediction, it reads the true next action from
the row at $t+1$. That value is never part of the agent's history-bounded input, so an agent-produced model cannot look ahead.

\paragraph{Simulation Validity (Task~2).}
For Task~2 the pipeline runs the submitted simulator on each held-out session across $100$
random seeds, which gives $900$ simulated trajectories for folds $1$ and $2$ and $1{,}000$
for fold~$3$. Every trajectory must obey the game's inventory rules, since a player can only
form a word the current letters allow. We keep a run only if at least $90\%$ of its
trajectories are valid, and we call such a run \emph{simulation-valid}. Four runs fall below
this bar and are dropped. The lowest run we keep is $90.2\%$ valid. Summary vectors and word
sequences are computed for each seed before pooling, so each seed is an independent draw, and
the unit of analysis is the run, not the seed.

\paragraph{Cost, Time, and Process.}
The pipeline records wall-clock time, token use, and monetary cost for every run. For
\texttt{claude-code} the cost comes from the provider's own accounting. For \texttt{codex} we
impute it from a blended rate of \$6.50 per $10^{6}$ tokens, which reflects a $94{:}6$ split
of input to output tokens priced at \$5 and \$30 per $10^{6}$ tokens
\citep{openai_api_pricing_2026}. The pipeline also logs the process traces used in the
analysis: how many candidate models and features the agent registered, how many modeling
decisions it recorded, and how long its submitted script was. Each agent also reports a
self-assessed score for its submission, which we use in the self-assessment analysis of
Section~\ref{sec:res-eda}.

\paragraph{Reproducibility.}
The agent, its model, and its reasoning-effort level are pinned for each run by the launch
script (for example, \texttt{claude-opus-4-7} is passed to Claude Code's model flag and the
reasoning level to its effort flag) and are stored with the run configuration. Every run's
provider, model, and effort are therefore reproducible from the released code, along with the
agent CLI versions reported in Section~\ref{sec:operators}.

\paragraph{A Note on Prompt Wording.}
The prompt inputs were not perfectly uniform across runs. The $18$ codex Task-1 full runs
used an earlier revision of two stable prompt components, the domain and constraints files,
while every other run used the current revision. The training data were identical, so this is
a difference in wording, not in information. Because the wording is fixed across effort levels
within these runs, it is orthogonal to the within-agent effort contrast that the main analysis
rests on, and so it does not confound the effort estimates. It does, however, line up exactly
with the full-versus-partial split for codex Task~1: every such full run carries the old
wording and every partial run the new one. We therefore do not read the codex Task-1
full-versus-partial difference as causal, and the full-information result of
Section~\ref{sec:res-eda}, which is specific to the $a$-KL condition, does not rely on it.

\section{Additional Exploratory Results}
\label{sec:app-extra}

This appendix collects descriptive material that fills in the picture of
Section~\ref{sec:res-eda} but is not needed for the main claims: the validity screen and
sample balance, the four excluded runs, how widely each agent searched, and which model
families the agents submitted.

\paragraph{Validity Screen and Balance.}
Of the $144$ scheduled runs, $140$ are simulation-valid and form the analysis sample. The
valid runs are balanced across agents ($70$ each) and across conditions ($36$ per condition,
and $32$ for the \texttt{a-DLD} condition, which loses two runs per agent), so the
comparisons in Section~\ref{sec:res-eda} are not driven by uneven counts.

\paragraph{The Four Excluded Runs.}
The four runs left out of the analysis show how agentic discovery breaks down. All four are
in the \texttt{a-DLD} condition and are the only runs below the $90\%$
simulation-validity threshold. Two are near-misses, valid in $885/1000$ and $806/900$
simulations, each failing on an out-of-support inventory draw at $t = 0$. The other two miss
the threshold by a wide margin: a codex submission valid in only $361/900$ simulations, whose
word-formation step skipped the set-rule constructibility check, and a claude-code submission
valid in just $1/900$ simulations, a maximum-effort run costing \$$15.37$ that treated
inventory as a set rather than a multiset. That last run is the most extreme instance of the
Task~2 self-assessment gap, since its own internal validation reported a strong score for a
model the evaluation pipeline almost entirely rejects. In all four cases the cause is a
specification error, an agent applying the wrong combinatorial rule, which the agent's own
checks did not flag.

\begin{figure}[htbp]\centering
\includegraphics[width=\textwidth]{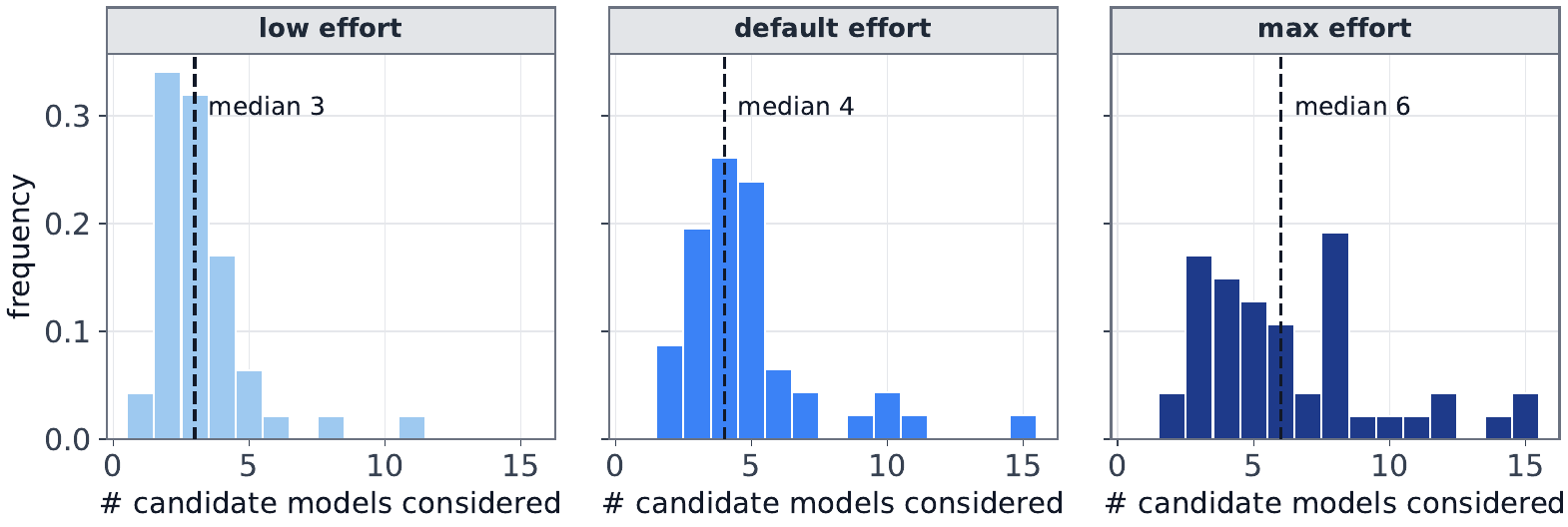}
\caption{Number of candidate models considered, one panel per reasoning-effort level. The
distribution shifts right as effort rises (median $3$, then $4$, then $6$).}
\label{fig:models}
\end{figure}

\paragraph{Search Width.}
Higher effort also widens the search. Figure~\ref{fig:models} shows relative-frequency
histograms of the number of candidate models an agent reports considering at each effort
level, with counts of $15$ or more pooled into the last bin and a dashed line at the median.
The distribution shifts to the right as effort rises, with the median going from $3$ at low
effort to $4$ at default and $6$ at maximum (means $3.2$, $5.3$, and $6.5$). This is the
descriptive counterpart of the registered-models process coordinate, which the multivariate
model of Section~\ref{sec:analysis} treats jointly with cost and performance.

\begin{figure}[htbp]\centering
\includegraphics[width=.9\textwidth]{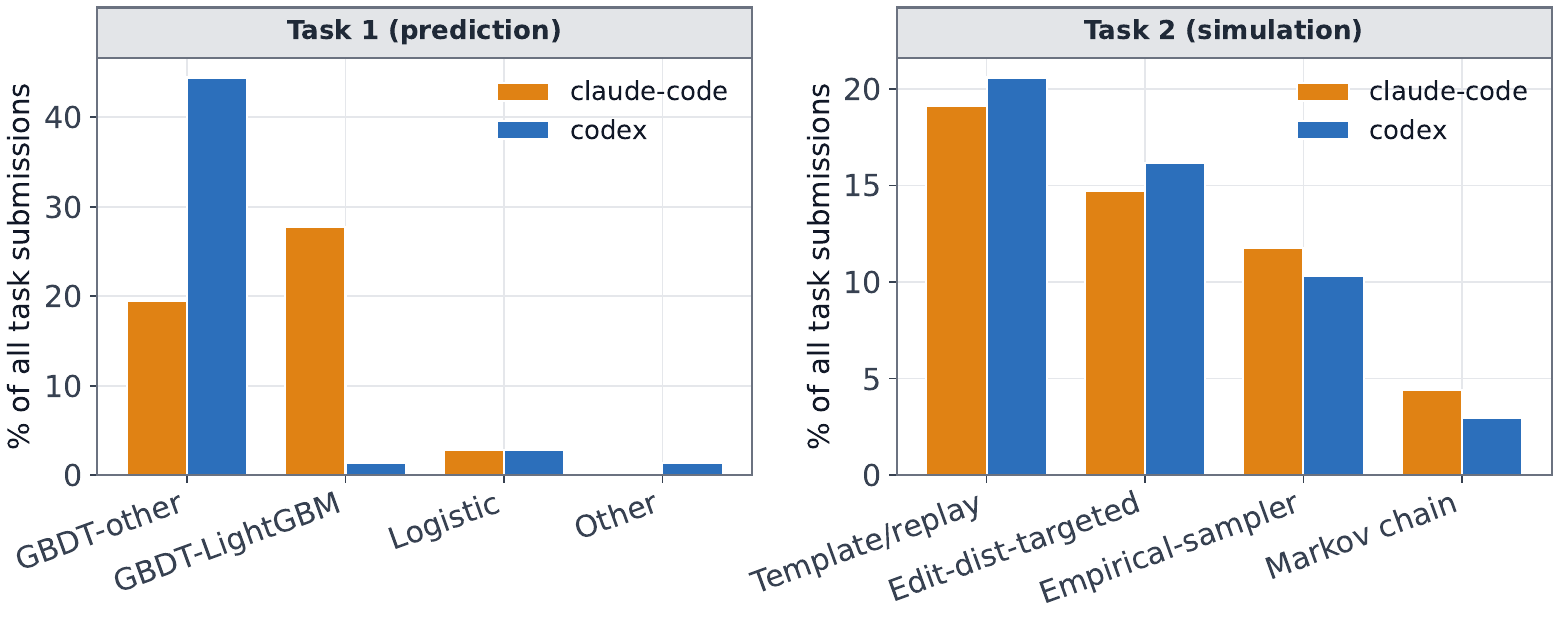}
\caption{Composition of submitted model families, by task, split by agent. Task~1 is almost
all gradient-boosted trees. Task~2 spreads across four simulator families.}
\label{fig:types}
\end{figure}

\paragraph{Model families.}
The submitted models show little variety on Task~1 and more on Task~2.
Figure~\ref{fig:types} gives the share of each task's submissions taken by every model or
simulator type, split by agent, with one panel per task. Predictive submissions are almost
all gradient-boosted trees ($93\%$, or $67$ of $72$), and the choice between scikit-learn's
HistGradientBoosting (shown as GBDT-other) and LightGBM follows the agent, with codex
preferring the first and claude-code the second, so on Task~1 the model type mostly stands in
for the agent. Simulation submissions divide among four families, with template/replay and
edit-distance-targeted constructions most common. Among these, the edit-distance-targeted construction, built for the distribution that $D_{\mathrm{LD}}$ scores, also attains the
lowest median $D_{\mathrm{LD}}$ of the four, about $0.16$ against $0.63$ for template/replay.

\end{document}